\documentclass[%
reprint,
longbibliography,
amsmath,amssymb,
amsfonts,
aps,
prx,
floatfix,
]{revtex4-2}
\usepackage{array}
\usepackage{autobreak}
\usepackage{bm}
\usepackage{booktabs}
\usepackage[dvipdfmx]{color}
\usepackage[dvipdfmx]{graphicx}
\usepackage{dcolumn}
\usepackage{tabularx}
\usepackage{float}
\usepackage{multirow}
\usepackage{newtxtext}
\usepackage{physics}
\usepackage{here}
\usepackage{mathtools}
\usepackage{comment}
\usepackage{url}

\newcommand{\dif}{\mathop{}\!\mathrm{d}}

\newcommand{\vct}[1]{\bm{#1}}      
\newcommand{\cvec}[1]{\vct{#1}}     
\newcommand{\mat}[1]{\mathbf{#1}}  
\begin{document}
\title{General spin models from noncollinear spin density functional theory and spin-cluster expansion}
\author{Tomonori Tanaka}
 \email{tanaka.t.da74@m.isct.ac.jp}
\author{Yoshihiro Gohda}
 \email{gohda@mct.isct.ac.jp}
\affiliation{Department of Materials Science and Engineering, Institute of Science Tokyo, Yokohama 226-8501, Japan}
\date{\today}
\begin{abstract}
We present a data-efficient framework for constructing general classical spin Hamiltonians by combining the spin-cluster expansion (SCE) with fully self-consistent noncollinear spin density functional theory (DFT).
The key idea is to fit the SCE model to magnetic torques rather than to total energies.
Because torques are site-resolved vectors, each spin configuration provides many informative regression targets, improving conditioning and substantially reducing the number of required DFT calculations, especially for large supercells.
Applied to the B20-type chiral magnets ${\rm Mn}_{1-x}{\rm Fe}_{x}{\rm Ge}$ and ${\rm Fe}_{1-y}{\rm Co}_{y}{\rm Ge}$, the resulting SCE models determine full pairwise exchange tensors---including isotropic exchange, symmetric anisotropic exchange, and the Dzyaloshinskii--Moriya interaction---and predict the helical spin period via a micromagnetic mapping.
The composition trends and the divergence of the period at the chirality sign-change point are well reproduced, in agreement with experiment.
Moreover, the systematic nature of SCE enables controlled assessment of interaction order: as the training spin configurations become more disordered, the lowest-order model loses torque accuracy, whereas including higher-order interactions restores predictive power.
These advances enable near-DFT-accurate spin models for finite-temperature magnetism and complex spin textures at modest computational cost, providing an extensible route to quantitative first-principles parameterization and predictive materials design.
An open-source implementation is available as a Julia package, \textit{Magesty.jl}.\end{abstract}
\maketitle
\clearpage
\section{Introduction}
\label{intro}
Recent advances in spintronic materials and technologies have been substantial, laying the groundwork for ultrafast, energy-efficient devices that exploit atomic-scale magnetic order and spin dynamics~\cite{Hirohata2020-fd, Bai2024-ib}.
In parallel, first-principles methods---especially density functional theory (DFT)~\cite{Hohenberg1964-yl, Kohn1965-fb}---have become central for the understanding and design of magnetic and spintronic materials.
However, fully \textit{ab initio} treatments often reach practical limits when addressing finite-temperature magnetism and mesoscopic spin dynamics.
Consequently, a broad class of statistical-mechanical models that coarse-grain the spin degrees of freedom has been developed~\cite{Skubic2008-kz, Yu-Lavrentiev2010-qx, Lavrentiev2011-dk, Evans2014-ng, Eriksson2017-lp, Yamada2019-vb, Rinaldi2024-wk}.
A natural coarse-graining is to map the electronic system onto a classical spin Hamiltonian that retains only the orientations of local magnetic moments as degrees of freedom.
The remaining electronic degrees of freedom, including charge and longitudinal-amplitude fluctuations, are thereby absorbed into effective interaction parameters that reproduce the energy of a given spin configuration.

In the immediate neighborhood of a locally stable magnetic configuration, a pairwise exchange-tensor model is often sufficient:
\begin{align}
\label{eq:lowest_order_hamiltonian}
E = \sum_{i,j}\sum_{\alpha,\beta} J_{ij}^{\alpha\beta}\, e_{i\alpha} e_{j\beta},
\end{align}
where the Greek indices $\alpha,\beta$ denote Cartesian components, $e_{i\alpha}$ is the $\alpha$ component of the unit vector $\hat{\vct{e}}_i$ specifying the moment direction at site $i$, and $J_{ij}^{\alpha\beta}$ is the two-body exchange interaction tensor between sites $i$ and $j$.
Such a lowest-order description, however, can become insufficient when one seeks a unified description over a broader set of spin configurations, in which case higher-order terms become necessary.
For example, temperature-dependent effective two-body isotropic terms (the Heisenberg interactions)---arising from the renormalization of higher-order interactions---can be important for modeling finite-temperature magnetic energetics~\cite{Tanaka2024-di}.
Furthermore, recent studies suggest that higher-order interactions have an important role for the stabilization of complex spin structures, such as chiral spin textures and skyrmions~\cite{Mendive-Tapia2021-yg, Gutzeit2021-av, Pan2024-ny}.
Accordingly, the importance of establishing quantitative methods to evaluate higher-order interactions is increasingly pronounced.

Systematically extending Eq.~(\ref{eq:lowest_order_hamiltonian}) to include higher-order terms in a Taylor-series fashion is challenging because the customary basis functions are not orthogonal.
This lack of orthogonality leads to ill-conditioned regressions and, under crystal symmetry, introduces cumbersome constraints among model parameters.
As a theoretically rigorous alternative, Drautz and F\"{a}hnle proposed the spin-cluster expansion (SCE) method~\cite{Drautz2004-id, Drautz2005-dg}.
Within SCE, the magnetic energy is a function of the unit vectors ${\hat{\vct{e}}_i}$ representing a spin direction at site $i$:
\begin{align}
\label{sce}
E_{\rm SCE}(\{{\hat{\vct{e}}_i}\}) = J_0 + \sum_{\mathcal{C}}\sum_{\vct l > 0}\sum_{\vct m} J_{\mathcal{C}{\vct l}{\vct m}}\Phi_{\mathcal{C}{\vct l}{\vct m}}(\{{\hat{\vct{e}}_i}\}),
\end{align}
where $J_0$ is a reference energy, $\mathcal{C}$ is the cluster index, $J_{\mathcal{C}{\vct l}{\vct m}}$ are the SCE coefficients, and $\Phi_{\mathcal{C}{\vct l}{\vct m}}$ is a basis function formed by products of spherical harmonics $Y_{lm}$,
\begin{align}
\begin{split}
\Phi_{\mathcal{C}{\vct l}{\vct m}}(\{{\hat{\vct{e}}_i}\}) &= (\sqrt{4\pi})^{n_{\mathcal{C}}} \prod_{i=1}^{n_{\mathcal{C}}} Y_{l_i m_i}(\hat{\vct{e}}_i),
\end{split}
\end{align}
with $n_{\mathcal{C}}$ the number of sites in cluster $\mathcal{C}$ and ${\vct l}=(l_1,\ldots,l_{n_{\mathcal{C}}})$, ${\vct m}=(m_1,\ldots,m_{n_{\mathcal{C}}})$.
Here ${\vct l>0}$ is understood componentwise, i.e., $l_i\ge 1$ for all sites in the cluster (not merely ${\vct l}\neq{\vct 0}$).
In spherical coordinates $(\theta,\phi)$,
\begin{align}
\label{sph}
Y_{lm}(\theta,\phi) = (-1)^m \sqrt{\cfrac{2l+1}{4\pi}\cfrac{(l-m)!}{(l+m)!}}P_{lm}(\cos\theta)e^{i m\phi},
\end{align}
where $P_{lm}$ are the associated Legendre polynomials related to the Legendre polynomials $P_l$ via
\begin{align}
P_{lm}(x) = (1-x^2)^{m/2}\cfrac{\dif^{m} P_l(x)}{\dif x^{m}}.
\end{align}
The set $\{{\Phi_{\mathcal C {\vct l} {\vct m}}}\}$, together with the constant term, forms a complete orthonormal basis set~\cite{Drautz2004-id}.
By increasing the maximum degree $l$ and the cluster size, the model can be systematically refined to better reproduce the underlying energy landscape.

In principle, the SCE coefficients $J_{\mathcal C {\vct l} {\vct m}}$ can be obtained by regressing to constrained noncollinear-spin DFT total energies, as shown in earlier SCE work~\cite{Singer2011-ou} and by analogy with the cluster expansion for substitutional alloys~\cite{Connolly1983-uk}.
In practice, however, magnetic interactions operate on energy scales far smaller than chemical interactions in alloys, which demands higher precision from DFT.
Moreover, whereas alloy variables are discrete, spin orientations are continuous, so the configuration space is vast; fits that rely only on total energies therefore require prohibitively many samples to reproduce the energy landscape with accuracy.
A practical solution is to augment the regression with information beyond total energies---namely, derivatives of the total energy with respect to spin orientations (magnetic torques).
The use of derivatives is well established elsewhere: in finite-displacement phonon calculations one fits DFT-computed atomic forces to build Taylor-expanded potentials~\cite{Parlinski1997-ob, Tadano2014-no}, and machine-learning interatomic potentials train on both energies and forces~\cite{Unke2021-ym}.
In magnetism, analogous strategies---fitting to magnetic torques from noncollinear-spin DFT---have been applied primarily to some spin models~\cite{Dederichs1984-jy, Brinker2019-fk, Jacobsson2022-nu, Rinaldi2024-wk}.
Integrating these ideas into the SCE framework yields a data-efficient and systematic route to the derivation of general spin models.

In this work, we present an efficient and scalable SCE framework fitting to magnetic torque data.
Unlike earlier SCE implementations~\cite{Singer2006-tb, Singer2011-ou, Dietermann2011-nq}, our approach is able to treat spin--orbit coupling (SOC) explicitly, which breaks global spin rotational invariance and necessitates anisotropic terms in spin models.
This enables nonperturbative evaluation of arbitrary anisotropic terms in the spin Hamiltonian, including single-ion anisotropy, symmetric exchange anisotropy, and the antisymmetric Dzyaloshinskii--Moriya interaction (DMI).
The paper is organized as follows.
Section~\ref{sec:theory} presents the theoretical framework: we formulate SCE using real spherical harmonics to ensure a real-valued energy in the presence of SOC (Sec.~\ref{ssec:real_sce}), construct symmetry-adapted basis functions to enforce crystal symmetries (Sec.~\ref{ssec:salc}), and describe the torque-fitting regression together with the Cartesian-polynomial representation of real spherical harmonics used to evaluate torque derivatives (Secs.~\ref{ssec:estimation} and~\ref{ssec:slm}).
Section~\ref{sec:application} applies the method to B20-type chiral magnets ${\rm Mn}_{1-x}{\rm Fe}_{x}{\rm Ge}$ and ${\rm Fe}_{1-y}{\rm Co}_{y}{\rm Ge}$.
After summarizing computational details (Sec.~\ref{ssec:details}) and our mean-field-based sampling scheme for generating temperature-controlled training spin configurations (Sec.~\ref{ssec:sampling}), we demonstrate the data efficiency of torque fitting (Sec.~\ref{ssec:data_efficiency}), derive micromagnetic parameters and the helical period (Sec.~\ref{ssec:micromagnetic}), and assess the role of higher-order interactions using more strongly disordered configurations (Sec.~\ref{higher-order}).
Finally, Sec.~\ref{sec:discussion} summarizes the main findings and discusses possible extensions.
Our implementation is released as an open-source Julia package, \textit{Magesty.jl}.

\section{Theory}
\label{sec:theory}
\subsection{Spin-cluster expansion with real spherical harmonics}
\label{ssec:real_sce}
In the original implementation of SCE~\cite{Drautz2004-id, Drautz2005-dg}, complex spherical harmonics were employed.
Subsequently, Ref.~\cite{Singer2006-tb} ensured a real-valued energy by assuming global spin rotation invariance (i.e., neglecting SOC) together with time-reversal symmetry.
When SOC is included, however, using complex spherical harmonics requires explicit constraints on the SCE coefficients to guarantee that the total energy remains real.
To avoid this bookkeeping and to simplify implementations, we work with real spherical harmonics.

The real spherical harmonics $Z_{lm}$ are defined by
\begin{equation}
Z_{lm} =
\begin{dcases}
\cfrac{(-1)^m}{\sqrt{2}}\!\left(Y_{lm} + (-1)^{m} Y_{l\bar m}\right), & m > 0,\\[4pt]
Y_{l0}, & m = 0,\\[4pt]
\cfrac{(-1)^{m}}{i\sqrt{2}}\!\left(Y_{l|m|} - (-1)^{|m|} Y_{l\overline{|m|}}\right), & m < 0,
\end{dcases}
\label{transform}
\end{equation}
where $\bar m \equiv -m$ and $\overline{|m|} \equiv -|m|$.
Collecting the harmonics for a fixed $l$ into column vectors ordered by increasing $m$ as
\(
\cvec{Z}_l \equiv (Z_{l\bar{l}},\dots,Z_{ll})^{\mathsf{T}}
\)
and
\(
\cvec{Y}_l \equiv (Y_{l\bar{l}},\dots,Y_{ll})^{\mathsf{T}},
\)
(with $(\cdot)^{\mathsf T}$ denoting transpose), the transformation in Eq.~(\ref{transform}) can be written compactly as
\begin{align}
\cvec{Z}_l = \mat{C}_{l}\,\cvec{Y}_l,
\end{align}
where $\mat{C}_{l}$ is a unitary matrix. An explicit form is
\begin{align}
\label{eq:Cl}
\mat{C}_{l} = \frac{1}{\sqrt{2}}
\begin{pmatrix}
i & 0  & \cdots & 0 & \cdots & 0  & -i(-1)^{l} \\
0  & i & \cdots & 0 & \cdots & -i(-1)^{l-1}  & 0 \\
\vdots & \vdots & \ddots & \vdots & \reflectbox{$\ddots$} & \vdots & \vdots \\
0  & 0  & \cdots & \sqrt{2} & \cdots & 0 & 0\\
\vdots & \vdots & \reflectbox{$\ddots$} & \vdots & \ddots & \vdots & \vdots \\
0 & 1 & \cdots & 0 & \cdots & (-1)^{l-1} & 0 \\
1 & 0 & \cdots & 0 & \cdots & 0 & (-1)^{l}
\end{pmatrix}\!,
\end{align}
with construction rules given in Ref.~\cite{Blanco1997-zs}.

The SCE written in terms of real spherical harmonics is obtained by replacing the complex-harmonic basis
$\Phi_{\mathcal C \vct l \vct m}$ in Eq.~(\ref{sce}) as
\begin{align}
\label{eq:real_basis}
E_{\rm SCE}(\{{\hat{\vct{e}}_i}\}) &= J_0 + \sum_{\mathcal{C}}\sum_{\vct l > 0}\sum_{\vct m} J_{\mathcal{C}{\vct l}{\vct m}}\Psi_{\mathcal{C}{\vct l}{\vct m}}(\{{\hat{\vct{e}}_i}\}), \\
\Psi_{\mathcal C \vct l \vct m}(\{\hat{\vct{e}}_i\})
&= (\sqrt{4\pi})^{n_{\mathcal C}} \prod_{i=1}^{n_{\mathcal C}} Z_{l_i m_i}(\hat{\vct{e}}_i).
\end{align}
The prefactor $(\sqrt{4\pi})^{n_\mathcal C}$ is chosen such that
the cluster basis functions are orthonormal with respect to the
scalar product
\begin{align}
\begin{split}
\braket{\Psi_{\mathcal C\vct l\vct m}}{\Psi_{\mathcal C'\vct l'\vct m'}}
&\equiv
\frac{1}{(4\pi)^N}
\int\dif\hat{\vct e}_1\cdots\int\dif\hat{\vct e}_N\;
\Psi_{\mathcal C\vct l\vct m}\,
\Psi_{\mathcal C'\vct l'\vct m'}\\
&=\delta_{\mathcal C\mathcal C'}\,
\delta_{\vct l\vct l'}\,
\delta_{\vct m\vct m'},
\end{split}
\end{align}
where $N$ is the total number of sites and $\int\dif\hat{\vct e}_i$ denotes integration over the unit sphere~\cite{Drautz2004-id, Singer2006-tb}.

For products over multiple sites, it is convenient to use Kronecker products.
Define the block vectors
\begin{align}
\cvec{Z}_{\vct l} \equiv \bigotimes_{i=1}^{n_{\mathcal C}} \cvec{Z}_{l_i},
\qquad
\cvec{Y}_{\vct l} \equiv \bigotimes_{i=1}^{n_{\mathcal C}} \cvec{Y}_{l_i},
\end{align}
and the corresponding transformation
\begin{align}
\label{eq:C_kron}
\mat{C}_{\vct l} \equiv \bigotimes_{i=1}^{n_{\mathcal C}} \mat{C}_{l_i}
= \mat{C}_{l_1} \otimes \mat{C}_{l_2} \otimes \cdots \otimes \mat{C}_{l_{n_{\mathcal C}}}.
\end{align}
Then the complex-to-real mapping factorizes as
\begin{align}
\label{eq:Z_kron}
\cvec{Z}_{\vct l} = \mat{C}_{\vct l} \cvec{Y}_{\vct l}.
\end{align}

\subsection{Construction of symmetry-adapted basis functions}
\label{ssec:salc}
A system that possesses a symmetry represented by a group $G$ must have an energy that is invariant under every symmetry operation in $G$.
In the framework of the SCE, this invariance can be achieved by constructing new basis functions as appropriate linear combinations of original ones ($\Psi_{\mathcal C\vct l\vct m}$), such that they form a basis for the identity representation of the group $G$.
For this purpose, we apply a group-theoretical projection-operator method.
The essential part of constructing the projection-operator method in the context of the SCE has already been explained in Ref.~\cite{Dietermann2011-nq}.
The key difference from the approach in Ref.~\cite{Dietermann2011-nq} is that while they assume global spin-rotational symmetry and time-reversal symmetry to ensure real-valuedness of the model, we do not impose the former symmetry.
Instead, we employ real spherical harmonics to ensure that the SCE model remains real-valued; although either set of spherical harmonics is equivalent in principle, using the real form offers the practical advantage that both the basis functions and the SCE coefficients remain strictly real in the implementation.
Below, we describe the projection-operator approach for the case using real spherical harmonics.

\begin{figure}
\centering
\includegraphics[width=0.9\columnwidth]{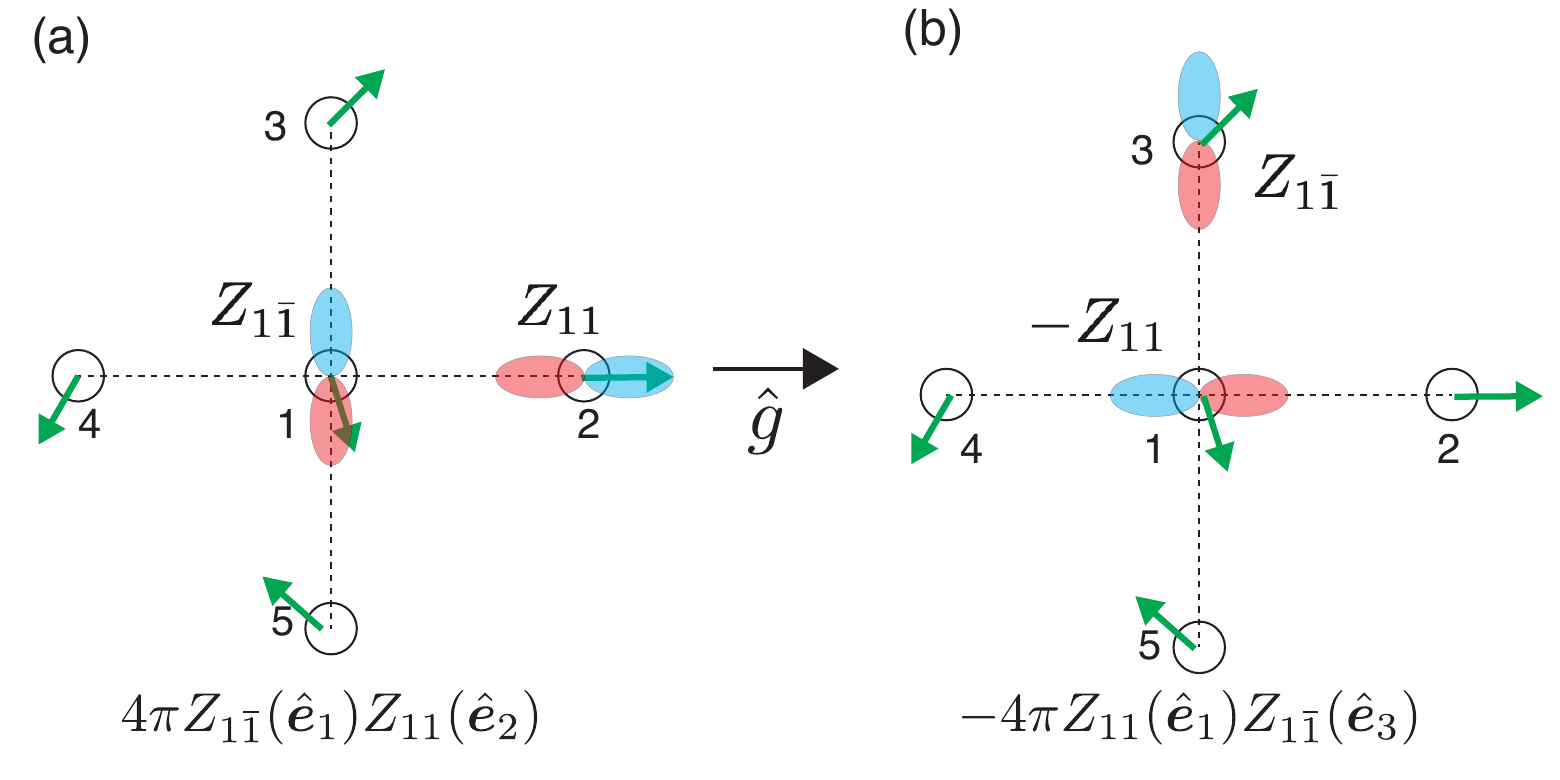}
\caption{
Example of a symmetry operation $\hat{g}$ acting on a basis function, where $\hat{g}$ is a counterclockwise rotation of $90^\circ$.
The circles and accompanying arrows represent atoms and spin directions, respectively.
(a) Original function $\Psi_{\mathcal{C}{\vct l}{\vct m}} = 4\pi Z_{1\bar{1}}(\hat{\vct e}_1)Z_{11}(\hat{\vct e}_2)$.
(b) Transformed function $\hat{g} \Psi_{\mathcal{C}{\vct l}{\vct m}} = -4\pi Z_{11}(\hat{\vct e}_1)Z_{1\bar{1}}(\hat{\vct e}_3)$.
For clarity, color images of $p_x$ (corresponding to $Z_{11}$) and $p_y$ (corresponding to $Z_{1\bar{1}}$) orbitals are also shown.
}
\label{sym_example}
\end{figure}

We consider a paramagnetic (gray) group $\bar{G}_0$ which consists of a space group $G_0$ and the time-reversal operation $\hat{\mathcal{T}}$ as 
\begin{align}
\bar{G}_0 = G_0 + \hat{\mathcal{T}} G_0,
\end{align}
where $\hat{\mathcal{T}}G_0$ is the set formed by the product of the time-reversal operation $\hat{\mathcal{T}}$ and the symmetry operations belonging to the space group $G_0$.
Note that we do not impose any symmetry of a specific magnetic order; magnetic space-group symmetries are not considered.
Magnetic space groups may emerge when the original spin Hamiltonian is approximated by an effective field Hamiltonian (e.g., by the mean-field approximation), but they do not represent the fundamental symmetry group that the original spin model is required to obey.
The projection operator of the identity irreducible representation $\hat{P}_{\rm id}$ is defined as follows:
\begin{align}
\label{projection}
\hat{P}_{\rm id} = \cfrac{1}{|\bar{G}_0|}\left(\sum_{\hat{g} \in G_0}\hat{g} + \sum_{\hat{g} \in G_0} \hat{\mathcal{T}} \hat{g}\right),
\end{align}
where $\hat{g}$ is the crystal-symmetry operation in $G_0$, and $|\bar{G}_0|$ is the order of group $\bar{G}_0$.
The desired symmetry-adapted basis functions are eigenfunctions of the matrix represented as
\begin{align}
	\bra{\Psi_{\mathcal{C}'{\vct l}'{\vct m}'}} \hat{P}_{\rm id} \ket{\Psi_{\mathcal{C}{\vct l}{\vct m}}}.
\end{align}
We decompose $\hat g$ into two commuting operations:
(i) a permutation of site labels $\hat g_{\rm site}$ induced by the mapping of atomic positions, and
(ii) a basis rotation $\hat g_{\rm rot}$ acting on spherical harmonics at fixed arguments (i.e., we rotate the \emph{basis functions} rather than the arguments $\hat{\boldsymbol e}_i$).
An example of $\hat g$ acting on a basis function $\Psi_{\mathcal C \boldsymbol l \boldsymbol m}$ is shown in Figs.~\ref{sym_example}(a) and (b).
From this example, the effect of the site mapping is clear; it merely permutes the site indices in the basis functions.
Accordingly, in what follows we focus on how the point-symmetry (basis-rotation) part $\hat g_{\rm rot}$ acts on the basis functions, and we omit the explicit site-permutation part $\hat g_{\rm site}$ for conciseness.
The matrix form of the first term in the right-hand side of Eq.~(\ref{projection}) is written as
\begin{align}
\begin{split}
	\label{eq:P_id_matrix_first}
	\cfrac{1}{|\bar{G}_0|}\sum_{\hat{g} \in G_0}\hat{g} &\doteq \cfrac{1}{|\bar{G}_0|} \sum_{\hat{g} \in G_0}\bra{\Psi_{\mathcal{C^{'}}{\vct l}^{'}{\vct m}^{'}}}\hat{g} \ket{\Psi_{\mathcal{C}{\vct l}{\vct m}}} \\
	&= \cfrac{1}{|\bar{G}_0|} \sum_{\hat{g} \in G_0} {\bm \Delta}_{\vct l}(\hat{g}).
\end{split}
\end{align}
Since the basis functions consist of the Kronecker product as Eq.~(\ref{eq:Z_kron}), ${\bm \Delta}_{\vct l}(\hat{g})$ is written as
\begin{align}
	\label{eq:rotmat_kron}
	\vct{\Delta}_{\vct l}(\hat{g}) = \bigotimes_{i=1}^{n_{\mathcal{C}}}{\Delta}_{l_i}(\hat{g}),
\end{align}
where ${\Delta}_{l_i}(\hat{g})$ is the matrix representation of the symmetry operation $\hat{g}$ acting on $\vct{Z}_{l_i}$.
If the point-symmetry operation part of $\hat{g}$ corresponds to a proper rotation $\hat{R}$, i.e., the determinant of the transform matrix equals $1$, ${\Delta}_{l}(\hat{g})$ can be obtained from the Wigner $D$-matrix and $\mat{C}_{\,l}$ in Eq.~(\ref{eq:Cl}) as follows~\cite{Blanco1997-zs}:
\begin{align}
	\begin{split}
	\Delta_{lmm'}(\hat{R}) &= \bra{Z_{lm}}\hat{R}\ket{Z_{lm'}}\\
	&= \sum_{m''m'''}C_{lmm''}^{*}C_{lm'm'''}\bra{Y_{lm''}}\hat{R}\ket{Y_{lm'''}}\\
	&= \sum_{m''m'''}C_{lmm''}^{*}D_{lm''m'''}(\hat{R})C_{lm'm'''}
	\end{split}
\end{align}
Therefore, ${\bm \Delta}_{\vct l}$ can be expressed as
\begin{align}
	\label{eq:matrix_form}
{\bm \Delta}_{\vct l}(\hat{R}) = {\bf C}_{\vct l}^{*}{\bf D}_{\vct l}(\hat{R}){\bf C}_{\vct l}^{\mathsf{T}},
\end{align}
where ${\bf D}_{\vct l}(\hat{R})$ is the Kronecker product of the Wigner $D$-matrices in the same manner as in Eq.~(\ref{eq:rotmat_kron}).
Even for improper rotations $\hat R'$ (with determinant $-1$), the matrix representations are obtained by noting that any $\hat R'$ can be written as the product of a corresponding proper rotation $\hat R$ and spatial inversion $\hat I$, i.e., $\hat R'=\hat R\hat I=\hat I\hat R$.
Because spin is an axial (pseudovector) quantity, spatial inversion does not reverse its direction; it merely permutes the site (cluster) indices. Hence,
\begin{align}
\hat I\,\Psi_{\mathcal C \vct l \vct m} \;=\; \Psi_{\mathcal C' \vct l \vct m},
\end{align}
where $\mathcal C' \equiv \hat I(\mathcal C)$ denotes the image of the cluster under inversion.
Therefore, the matrix representation of any improper rotation $\hat R'=\hat R\hat I$ can be obtained in the same manner as for proper rotations (up to the index permutation induced by $\hat I$).
Finally, the time-reversal operator $\hat{\mathcal T}$ flips the spin direction, $\hat{\mathcal T}\,\hat{\vct{e}}_i \mapsto -\hat{\vct{e}}_i$.
Therefore,
\begin{align}
\hat{\mathcal T}\,\Psi_{\mathcal C \vct{l} \vct{m}}\big(\{\hat{\vct{e}}_i\}\big)
= (-1)^{\sum_{i=1}^{n_{\mathcal C}} \! l_i}\,
\Psi_{\mathcal C \vct{l} \vct{m}}\big(\{\hat{\vct{e}}_i\}\big).
\end{align}
It follows that only cluster terms with an even $\sum_{i=1}^{n_{\mathcal C}} l_i$ are $\hat{\mathcal T}$-even and therefore allowed as a time-reversal-invariant function, whereas terms with an odd $\sum_{i=1}^{n_{\mathcal C}} l_i$ are $\hat{\mathcal T}$-odd and thus excluded~\cite{Singer2006-tb}.
Combining this with the spatial symmetry operations, we can construct the projector onto the identity irreducible representation.
The symmetry-adapted basis functions are the eigenfunctions of $\hat{P}_{\rm id}$ with eigenvalue $1$~\cite{Blanco1997-zs}.
In the actual calculations, we used Spglib library for the evaluation of crystal symmetry~\cite{Togo2024-vg}.

\subsection{Estimation of SCE coefficients utilizing local magnetic torques}
\label{ssec:estimation}
We estimate the SCE coefficients using effective magnetic torques obtained from constrained noncollinear-spin DFT~\cite{Ma2015-ck}.
The torque at atomic site \(i\), \(\boldsymbol T_i\), is defined as the energy gradient on the unit sphere cast into a physical torque,
\begin{align}
\label{eq:torque_dft}
\boldsymbol T_{{\rm DFT}, i}
\equiv
\hat{\boldsymbol e}_i \times \!\left(-\,\nabla_{\hat{\boldsymbol e}_i} E_{\rm DFT}\right)
\;=\; \boldsymbol {\mathfrak m}_i \times \boldsymbol B_i,
\end{align}
where \(E_{\rm DFT}\) is the total energy, \({\mathfrak m}_i=|{\mathfrak m}_i|\hat{\boldsymbol e}_i\) is the local magnetic moment, and \(\boldsymbol B_i\) is a site-averaged effective magnetic field. At self-consistency, \(\boldsymbol B_i\) is transverse to \(\hat{\boldsymbol e}_i\) and equal in magnitude and opposite in direction to the constraining field~\cite{Ma2015-ck}.
Hence both sides of \(-\nabla_{\hat{\boldsymbol e}_i} E_{\rm DFT}=|{\mathfrak m}_i| \boldsymbol B_i\) lie in the tangent plane at \(\hat{\boldsymbol e}_i\)~\cite{Brinker2019-fk}.
For completeness, the gradient on the unit sphere with respect to \(\hat{\boldsymbol e}_i=(e_{ix},e_{iy},e_{iz})\) is
\begin{align}
\nabla_{\hat{\boldsymbol e}_i} f(\hat{\boldsymbol e}_i)
= \nabla f(\hat{\boldsymbol e}_i) - \bigl(\hat{\boldsymbol e}_i\!\cdot\!\nabla f(\hat{\boldsymbol e}_i)\bigr)\,\hat{\boldsymbol e}_i, \label{eq:grad_e}\\
\nabla f(\hat{\boldsymbol e}_i)
= \frac{\partial f}{\partial e_{ix}}\,\boldsymbol{i}
 + \frac{\partial f}{\partial e_{iy}}\,\boldsymbol{j}
 + \frac{\partial f}{\partial e_{iz}}\,\boldsymbol{k}, \label{eq:grad}
\end{align}
with \(\boldsymbol{i},\boldsymbol{j},\boldsymbol{k}\) the Cartesian unit vectors.

Let \(\boldsymbol J\) be the \(N_{\rm I}\)-vector collecting the irreducible, symmetry-adapted SCE coefficients (excluding \(J_0\)).
The SCE energy can be written as
\begin{align}
E_{\rm SCE}(\{\hat{\boldsymbol e}_i\}) = J_0 + \boldsymbol u^{\mathsf T}\boldsymbol J,
\end{align}
where \(\boldsymbol u = \partial E_{\rm SCE}(\{\hat{\boldsymbol e}_i\})/\partial \boldsymbol J\) is a vector function of the spin directions.
The corresponding torque is
\begin{align}
\boldsymbol T_{{\rm SCE},i}
= \hat{\boldsymbol e}_i \times \bigl(-\nabla_{\hat{\boldsymbol e}_i} E_{\rm SCE}\bigr)
= \hat{\boldsymbol e}_i \times \Bigl[-\bigl(\nabla_{\hat{\boldsymbol e}_i}\boldsymbol u^{\mathsf T}\bigr)\,\boldsymbol J\Bigr].
\label{eq:t_i_sce}
\end{align}
Collecting all sites, this becomes the linear relation
\begin{align}
\label{eq:t_sce}
\boldsymbol T_{\rm SCE} \;=\; \mathbf A\,\boldsymbol J,
\end{align}
where \(\boldsymbol T_{\rm SCE}\in\mathbb{R}^{3N_{\rm a}}\) and \(\mathbf A\in\mathbb{R}^{3N_{\rm a}\times N_{\rm I}}\) has block rows
\begin{widetext}
\[
\mathbf A \;=\;
\begin{pmatrix}
-\hat{\boldsymbol e}_1 \times \bigl(\nabla_{\hat{\boldsymbol e}_1}u_1\bigr) &
-\hat{\boldsymbol e}_1 \times \bigl(\nabla_{\hat{\boldsymbol e}_1}u_2\bigr) &
\cdots &
-\hat{\boldsymbol e}_1 \times \bigl(\nabla_{\hat{\boldsymbol e}_1}u_{N_{\rm I}}\bigr) \\
-\hat{\boldsymbol e}_2 \times \bigl(\nabla_{\hat{\boldsymbol e}_2}u_1\bigr) &
-\hat{\boldsymbol e}_2 \times \bigl(\nabla_{\hat{\boldsymbol e}_2}u_2\bigr) &
\cdots &
-\hat{\boldsymbol e}_2 \times \bigl(\nabla_{\hat{\boldsymbol e}_2}u_{N_{\rm I}}\bigr) \\
\vdots & \vdots & \ddots & \vdots \\
-\hat{\boldsymbol e}_{N_{\rm a}} \times \bigl(\nabla_{\hat{\boldsymbol e}_{N_{\rm a}}}u_1\bigr) &
-\hat{\boldsymbol e}_{N_{\rm a}} \times \bigl(\nabla_{\hat{\boldsymbol e}_{N_{\rm a}}}u_2\bigr) &
\cdots &
-\hat{\boldsymbol e}_{N_{\rm a}} \times \bigl(\nabla_{\hat{\boldsymbol e}_{N_{\rm a}}}u_{N_{\rm I}}\bigr)
\end{pmatrix}\!.
\]
\end{widetext}
For \(N_{\rm data}\) spin configurations, we stack the site torques and design matrices:
\begin{align}
\widetilde{\boldsymbol T}_{\rm SCE} \;=\; \widetilde{\bf A}\,\boldsymbol J,\qquad
\widetilde{\boldsymbol T} \equiv 
\begin{pmatrix}
\boldsymbol T_{1}\\ \boldsymbol T_{2}\\ \vdots\\ \boldsymbol T_{N_{\rm data}}
\end{pmatrix},\quad
\widetilde{\bf A} \equiv 
\begin{pmatrix}
\mathbf A_{1}\\ \mathbf A_{2}\\ \vdots\\ \mathbf A_{N_{\rm data}}
\end{pmatrix}.
\end{align}
Identifying \(\widetilde{\boldsymbol T}_{\rm SCE}\) with the DFT torques \(\widetilde{\boldsymbol T}_{\rm DFT}\),
we determine \(\boldsymbol J\) by minimizing the squared Frobenius-norm loss
\begin{align}
\mathcal L \;=\; \bigl\|\,\widetilde{\boldsymbol T}_{\rm DFT} - \widetilde{\boldsymbol T}_{\rm SCE}\,\bigr\|_2^{\,2}.
\end{align}
When the system is well determined, the least-squares estimate of $\bm J$ is given by
$\,{\bm J}=(\widetilde{\bf A}^{\mathsf{T}}\widetilde{\bf A})^{-1}\widetilde{\bf A}^{\mathsf{T}}\widetilde{\boldsymbol T}_{\rm DFT}$.

\subsection{Real spherical harmonics represented as polynomials of Cartesian coordinates}
\label{ssec:slm}
In torque-fitting regression formalism, we must evaluate gradients of the spherical harmonics with respect to the unit vector, cf. Eq.~(\ref{eq:t_i_sce}).
Differentiation in spherical coordinates $(\theta,\phi)$ is numerically ill-conditioned near the poles ($\sin\theta\!\to\!0$).
To avoid these coordinate singularities, we represent the spherical harmonics as polynomials in the Cartesian components of the unit vector, which yields well-behaved analytic derivatives via $\partial/\partial e_x$, $\partial/\partial e_y$, and $\partial/\partial e_z$.
The gradient of the complex spherical harmonics with respect to the unit vector was already derived in Ref.~\cite{Drautz2020-ao}.
Here, we present expressions for the gradients of the real spherical harmonics in their Cartesian-polynomial form.
See also the Supplemental Material Sec. S1 for details of the derivations~\cite{SI}.

The complex spherical harmonics $Y_{lm}$ for $m \ge 0$ can be written in Cartesian coordinates as follows~\cite{Drautz2020-ao}:
\begin{align}
	Y_{lm}(\hat{\vct{e}}_i) &= (e_{ix} + ie_{iy})^m\bar{P}_{lm}(e_{iz}),\label{eq:y_cart}\\
	\bar{P}_{lm}(e_{iz}) &= (-1)^m\sqrt{\cfrac{2l+1}{4\pi}\cdot\cfrac{(l-m)!}{(l+m)!}}\cfrac{\dif^m}{\dif e_{iz}^m}P_{l}(e_{iz}).\label{eq:bar_p_cart}
\end{align}
For $m<0$, the corresponding expressions follow from the relation $Y_{l\bar m} = (-1)^{|m|} Y_{l|m|}^{*}$.
Hereafter, we omit the arguments $\hat{\vct{e}}_i$ and the spin index $i$ when not explicitly required.
From Eqs.~(\ref{transform}), (\ref{eq:y_cart}), and (\ref{eq:bar_p_cart}) we obtain the real spherical harmonics in Cartesian coordinates as follows:
\begin{widetext}
\begin{align}
Z_{lm} =
\begin{dcases} (-1)^{m}\sqrt{2} \bar{P}_{lm}(e_{z}) \sum^{\left[ m/2 \right]}_{k=0} (-1)^{k} \binom{m}{2k}e_{x}^{m-2k}e_{y}^{2k}, & m > 0\\ 
\bar{P}_{l0}(e_z), & m = 0\\
(-1)^{m}\sqrt{2} \bar{P}_{l|m|}(e_{z}) \sum^{\left[(|m|-1)/2 \right]}_{k=0} (-1)^{k} \binom{|m|}{2k+1}e_{x}^{|m|-(2k+1)}e_{y}^{2k+1}, & m < 0
\end{dcases}
,
\end{align}
\end{widetext}
where $[x]$ denotes the floor function (i.e., the largest integer less than or equal to $x$), and $\binom{m}{k}$ is the binomial coefficient.

The gradient of the real spherical harmonics with respect to the unit vector $\hat{\vct{e}}$ can be derived using Eqs.~(\ref{eq:grad_e}) and (\ref{eq:grad}) as follows:
\begin{align}
\nabla_{\hat{\vct{e}}} Z_{lm}(\hat{\vct{e}}) &= \nabla Z_{lm}(\hat{\vct{e}}) - (\hat{\vct{e}} \cdot \nabla Z_{lm}(\hat{\vct{e}}))\hat{\vct{e}}, \label{eq:grad_e_sph}\\
\nabla Z_{lm}(\hat{\vct{e}}) &= \cfrac{\partial Z_{lm}}{\partial e_{x}}{\vct i} + \cfrac{\partial Z_{lm}}{\partial e_{y}}{\vct j} + \cfrac{\partial Z_{lm}}{\partial e_{z}}{\vct k}.
\end{align}
The partial derivatives ${\partial Z_{lm}}/{\partial e_\alpha}$ ($\alpha \in x, y, z$) for $m > 0$ are given by:
\begin{widetext}
\begin{align}
\cfrac{\partial Z_{lm}}{\partial e_x} &= (-1)^m \sqrt{2} m\bar{P}_{lm}(e_{z}) \sum^{[(m-1)/2]}_{k=0}(-1)^k \binom{m-1}{2k} e_{x}^{m-1-2k} e_{y}^{2k}, \\
\cfrac{\partial Z_{lm}}{\partial e_y} &= -\cfrac{\partial Z_{l\bar{m}}}{\partial e_x}, \\
\cfrac{\partial Z_{lm}}{\partial e_z} &= (-1)^m \sqrt{2} \left( \cfrac{\dif \bar{P}_{l m}(e_z)}{\dif e_z} \right)
\sum_{k=0}^{[m/2]} (-1)^k \binom{m}{2k} e_x^{m - 2k} e_y^{2k}.
\end{align}
For $m = 0$,
\begin{align}
	\cfrac{\partial Z_{l0}}{\partial e_x} &= \cfrac{\partial Z_{l0}}{\partial e_y} = 0, \\
	\cfrac{\partial Z_{l0}}{\partial e_z} &= \cfrac{\dif\bar{P}_{l 0}(e_z)}{\dif e_z}.
\end{align}
For $m < 0$,
\begin{align}
\cfrac{\partial Z_{lm}}{\partial e_x}& =
(-1)^m \sqrt{2} |m| \bar{P}_{l|m|}(e_{z}) \sum^{[(|m|-2)/2]}_{k=0} (-1)^k \binom{|m|-1}{2k+1} e_{x}^{|m|-2-2k} e_{y}^{2k+1}, \label{z_ex}\\
\cfrac{\partial Z_{lm}}{\partial e_y} &= \cfrac{\partial Z_{l|m|}}{\partial e_x},\\
\cfrac{\partial Z_{lm}}{\partial e_z} &=(-1)^m \sqrt{2} \left( \cfrac{\dif \bar{P}_{l|m|}(e_z)}{\dif e_z} \right)
\sum_{k=0}^{[(|m|-1)/2]} (-1)^k \binom{|m|}{2k+1} e_x^{|m| - 2k - 1} e_y^{2k + 1}.
\end{align}
\end{widetext}
We adopt the convention that whenever the upper limit of a summation is smaller than the lower limit, the sum is taken to be empty and hence evaluates to zero.
In particular, in Eq.~(\ref{z_ex}) the upper limit $\big\lfloor(|m|-2)/2\big\rfloor$ is negative for $|m|=1$, so the sum vanishes and $\partial Z_{l\bar{1}}/\partial e_x=0$.
Based on the above expressions, we can evaluate the gradients of the SCE basis functions for a given spin configuration.

\section{Application}
\label{sec:application}
B20-type chiral magnetic compounds ${\rm Mn}_{1-x}{\rm Fe}_{x}{\rm Ge}$ and ${\rm Fe}_{1-y}{\rm Co}_{y}{\rm Ge}$ exhibit helical spin structures~\cite{Ludgren1970-mk, Lebech1989-eg, Kanazawa2011-ks, Yu2011-ni, Grigoriev2013-vb, Shibata2013-tg, Kanazawa2016-hf, Siegfried2017-qq, Spencer2018-mh, Turgut2018-ep, Guang2024-nc}.
These compounds have a moderate SOC strength, making them suitable testbeds for developing DFT-based methods to evaluate the DMI~
\cite{Gayles2015-ow, Kikuchi2016-iu, Koretsune2018-jh, Grytsiuk2019-pa}.
Here, we apply the SCE method to the B20-type compounds and examine the applicability via the evaluation of exchange interactions, DMI, and helical spin period.

\subsection{Computational details}
\label{ssec:details}
For the constrained noncollinear-spin DFT calculations, we used the VASP package~\cite{Kresse1996-oa, Kresse1999-sm} with the projector augmented-wave method~\cite{Blochl1994-fq} and the generalized-gradient approximation functional of Perdew, Burke, and Ernzerhof~\cite{Perdew1996-sf}.
The constrained local moment approach~\cite{Ma2015-ck} was employed to fix the directions of the atomic magnetic moments, while allowing their magnitudes to relax.
Spin--orbit coupling was included self-consistently.
The electronic energy-convergence threshold was set to $10^{-7}$~eV/atom.
A plane-wave cutoff energy of $300$~eV and a $5\times5\times5$ $k$-point grid were used for a $2\times2\times2$ cubic supercell (64 atoms).
The lattice constant and internal atomic positions were taken from the experimental values of FeGe reported in Ref.~\cite{Lebech1989-eg}.
Atomic radii of $1.242$~$\mathring{\text{A}}$\ for Fe and $1.161$~$\mathring{\text{A}}$\ for Ge were used for the constrained local moment calculations.
We used a 2$\times$2$\times$2 cubic supercell including 64 atoms with experimental structural parameters of FeGe~\cite{Lebech1989-eg} and constructed an SCE model using spin configurations sampled from states close to the ferromagnetic state, following the sampling procedure described in Sec.~\ref{ssec:sampling}.
For the composition-dependent micromagnetic parameters discussed in Sec.~\ref{ssec:micromagnetic}, we emulated changes in chemical composition by electron and hole doping, while keeping the structural parameters fixed to those of FeGe.
Note that the SCE coefficients obtained from a finite supercell are subject to finite-size effects.
In a finite supercell with periodic boundary conditions, these parameters do not represent interactions between isolated individual sites, but rather effective interactions defined for sites $i$ and $j$ under periodic repetition.
These finite-size effects decrease as the supercell size increases.

We also comment on the relaxation of the moment magnitudes in constrained noncollinear spin DFT calculations.
In classical spin models, including the SCE model, ${\hat{\vct{e}}}_i$ is not necessarily interpreted as a local magnetic moment of fixed magnitude, but rather as a dimensionless unit vector representing the orientation of the local magnetic moment.
From this perspective, just as the total energy is treated as a function of ${\hat{\vct{e}}}_i$, the magnitude of the local magnetic moment may also be regarded as a dependent variable determined by $\{{\hat{\vct{e}}}_i\}$~\cite{Drautz2004-id}.
The energy change arising from variations in the moment magnitudes can then be effectively incorporated into SCE coefficients.

\subsection{Sampling method based on the mean-field approximation}
\label{ssec:sampling}
Constructing an SCE model requires sampling spin configurations that adequately span the magnetic states of interest.
In general, building a model that reproduces energies across arbitrary magnetic states with high accuracy within a realistic computational budget will demand  sophisticated strategies, which we defer to future work.
In the applications presented below, we instead assume that the relevant magnetic states are known \textit{a priori}.
For this setting, we employ a practical mean-field-based sampling scheme for the Heisenberg model that efficiently generates spin configurations representative of the target states~\cite{Tanaka2024-di,Gyorffy1985-gl,Mendive-Tapia2022-wq}.
Within the mean-field Heisenberg model, the single-site orientational probability density $P(\hat{\vct e}_i)$ takes the von Mises--Fisher form,
\begin{align}
	P(\hat{\vct{e}}_i) = \cfrac{3{\mathrm m}/\tau}{4\pi \sinh(3{\mathrm m}/\tau)} \exp\left( \cfrac{3{\mathbf m} \cdot \hat{\vct{e}}_i}{\tau} \right),
	\label{eq:mfa_sampling}
\end{align}
where ${\mathbf m} = \langle\hat{\vct e}\rangle = {\mathrm m}\hat{\vct e}$ ($0 \leq {\mathrm m} \leq 1$) is the site-independent thermal average of the spin and $\tau = T/T_{\rm c}^{\rm MFA}$ is a reduced temperature ($T$ is the temperature and $T_{\rm c}^{\rm MFA}$ is the mean-field critical temperature).
The parameters ${\mathrm m}$ and $\tau$ are self-consistently related; fixing one automatically determines the other.
Therefore, by choosing $\tau$, one can sample $\hat{\vct e}_i$ from this distribution to generate spin configurations representative of the target magnetic disorder within the mean-field approximation.
Please refer to Ref.~\cite{Tanaka2024-di} for the derivation and further details.
In this study, we used $\tau = 0.1$ for demonstrating data efficiency (Sec.~\ref{ssec:data_efficiency}) and for deriving micromagnetic parameters (Sec.~\ref{ssec:micromagnetic}), while $\tau = 0.3$ was additionally employed to evaluate higher-order interaction terms (Sec.~\ref{higher-order}).
All constrained noncollinear spin DFT calculations for the spin configurations generated according to the sampling procedure described above converged successfully.

\subsection{Spin-cluster expansion and micromagnetic models}
For the SCE model, we first consider the constant term and two-body interactions with the lowest $l$-index as
\begin{align}
\begin{split}
E_{\rm SCE} = 
J_0 + \sum_{i, j}\sum_{l_i, l_j = 1}^{l_{\rm max} = 1} \sum_{m_i= -l_i}^{l_i} \sum_{m_j= -l_j}^{l_j} J_{ij\vct{l}\vct{m}} Z_{l_i m_i}(\hat{\vct{e}}_i) Z_{l_j m_j}(\hat{\vct{e}}_j).
\end{split}
\label{eq:SCE_fege}
\end{align}
The sum $\sum_{i,j}$ runs over all ordered pairs with $i \neq j$ (self-terms are excluded), and each pair is counted twice.
We included all possible two-body pairs (Fe--Fe, Fe--Ge, and Ge--Ge).
While the Ge moment is largely induced and therefore small, including Ge-related pairs avoids imposing an \textit{a priori} truncation and lets their contributions be assessed quantitatively.
By constructing symmetry-adapted basis functions, the number of independent SCE coefficients excluding the constant term resulted in 196.
This SCE model is equivalent to the generalized Heisenberg model consisting of the isotropic exchange term, DMI, and anisotropic symmetric interaction term as
\begin{align}
	\begin{split}
	E = &-\sum_{i, j} J_{ij} \hat{\vct{e}}_i \cdot \hat{\vct e}_j \\
	&+ \sum_{i, j} {\vct D}_{ij} \cdot(\hat{\vct{e}}_i \times \hat{\vct e}_j)\\
	&+ \sum_{i, j} {\hat{\vct{e}}_i}^{\mathsf{T}} {\vct \Gamma}_{ij} \hat{\vct e}_j,
	\end{split}
	\label{eq:fege_spin_model}
\end{align}
where $J_{ij}$ is the isotropic exchange parameter, ${\vct D}_{ij}$ is the DMI vector, and ${\vct \Gamma}_{ij}$ is the anisotropic symmetric interaction matrix with conditions ${\vct \Gamma}_{ij} = {\vct \Gamma}_{ij}^{\mathsf{T}}$ and $\tr[{\vct \Gamma}_{ij}] = 0$.
The conversion rule from SCE coefficients to $J_{ij}$, ${\vct D}_{ij}$, and ${\vct \Gamma}_{ij}$ is explained in the Appendix.

We make a few remarks on the spin-model parameters.
In our formalism, symmetry-adapted on-site (one-body) anisotropic terms can also be included.
Although the B20-type structure considered here is macroscopically cubic, the local symmetry at each atomic site is lower than cubic, so a uniaxial anisotropy term, corresponding to the $l = 2$ spherical-harmonic components, can in principle appear.
We therefore performed additional fittings including this on-site term for all compositions.
However, we confirmed that both the on-site term itself and its inclusion have only negligible effects on the other SCE coefficients and the resulting micromagnetic parameters $A$, $D$, and $\lambda$.
We therefore neglected the on-site term in the present study.
In addition, the symmetry-adapted basis functions are automatically consistent with Moriya rules~\cite{Moriya1960-ib}.
For example, in B20-type FeGe, the DMI vectors for the Fe--Fe and Ge--Ge bonds oriented along the [111] direction are constrained by the threefold rotation symmetry to be parallel or antiparallel to the rotation axis.

Following Refs.~\cite{Kikuchi2016-iu, Koretsune2018-jh, Grytsiuk2019-pa, Bornemann2019-vg}, we employ a micromagnetic energy functional derived from Eq.~(\ref{eq:fege_spin_model}) together with the B20 crystal geometry.
Exploiting the cubic symmetry of the B20 structure, the energy functional can be written as~\cite{Bornemann2019-vg}
\begin{align}
\begin{split}
		E[\vct{m}(\vct r)]
&= \cfrac{1}{V}\int_{V} \dif{\vct r}\,\left[
  A\!\left( (\nabla m_x)^2 + (\nabla m_y)^2 + (\nabla m_z)^2 \right)
\right.\\
&\qquad \left. {}+ D\,{\vct m}\cdot\left(\nabla\times{\vct m}\right) \right],
\end{split}
\end{align}
where $\vct m(\vct r) = \big(m_x(\vct r), m_y(\vct r), m_z(\vct r)\big)$ is the normalized magnetization field, $A$ is the spin-stiffness constant, and $D$ is the spiralization constant; under cubic symmetry, the anisotropic symmetric exchange term does not contribute to the micromagnetic parameters.
The parameters $A$ and $D$ are calculated as~\cite{Grytsiuk2019-pa}
\begin{align}
A &= \cfrac{1}{2}\sum_{i, j} J_{ij} {\vct R}_{ij} \cdot {\vct R}_{ij}, \label{eq:stiffness}\\
D &= \sum_{i, j} {\vct D}_{ij} \cdot{\vct R}_{ij}, \label{eq:spiralization}
\end{align}
where ${\vct R}_{ij} = {\vct R}_{j} - {\vct R}_{i}$ is the relative vector between atom $i$ and $j$.
The helical period $\lambda$ then follows as
\begin{align}
\lambda = 4\pi\left| A/D\right|. 
\label{eq:lambda}
\end{align}
The sign of $D$ does not affect $\lambda$ but sets the handedness (right- or left-handed) of the helix~\cite{Grigoriev2013-vb}. 
In the exchange-DMI-only model, $\lambda$ diverges as $D \to 0$, corresponding to the ferromagnetic limit.

Note that the evaluation of the micromagnetic parameters $A$ and $D$ in Eqs.~(\ref{eq:stiffness}) and (\ref{eq:spiralization}) explicitly depends on the intersite distance ${\vct R}_{ij}$.
This point is closely related to the finite-size effects discussed in Sec.~\ref{ssec:details}, since in a finite supercell the interactions are effectively renormalized under periodic boundary conditions.
Therefore, a $1\times1\times1$ cell is not sufficient for their proper evaluation, because the renormalized interactions in such a cell do not retain the distance dependence required in these expressions.

\subsection{Data efficiency of the regression scheme}
\label{ssec:data_efficiency}

\begin{figure}
	\includegraphics[width=95mm]{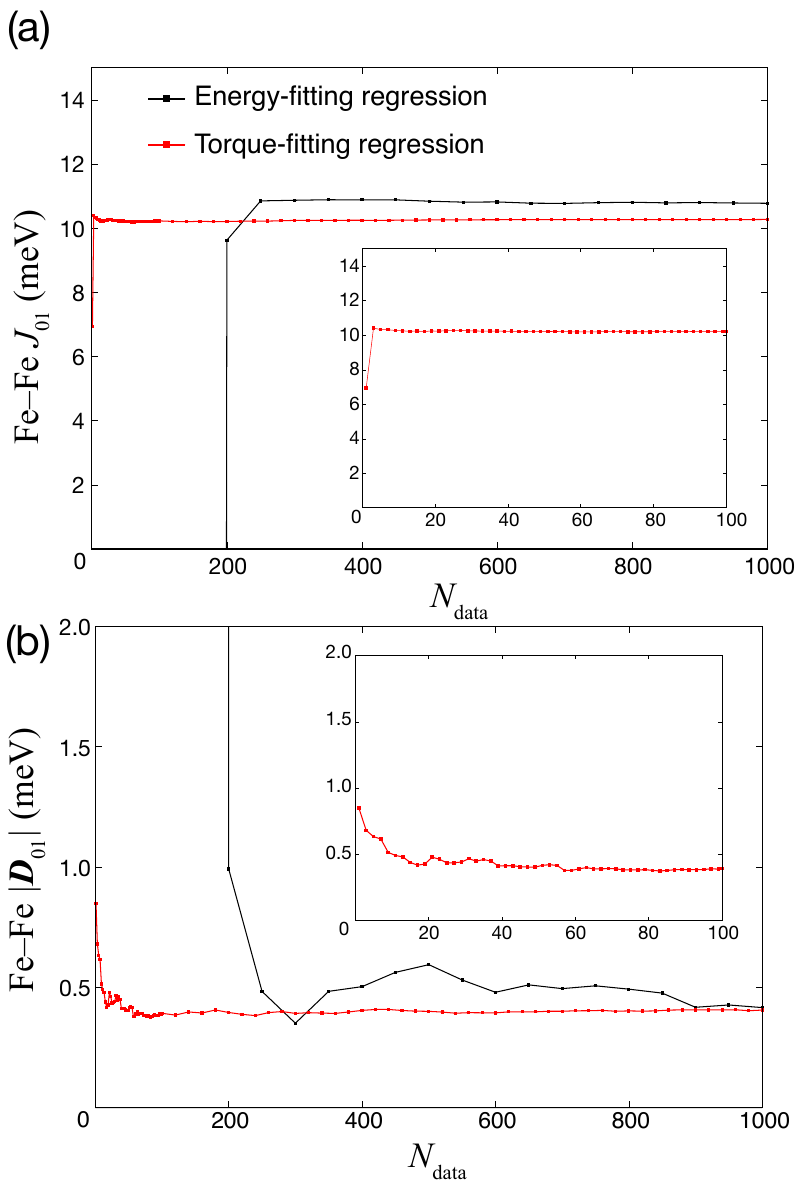}
	\caption{Comparison of the convergence with respect to the number of DFT data points $N_{\mathrm{data}}$ for energy-fitting (black) and torque-fitting (red) regression. 
Panels (a) and (b) show the first-nearest-neighbor Fe--Fe exchange coupling $J_{01}$ and the magnitude of the DMI vector $|{\vct D}_{01}|$, respectively. 
The insets enlarge the torque-fitting regression results for $N_{\mathrm{data}} \le 100$.}
\label{fig:ndata}
\end{figure}

We first demonstrate that torque-fitting regression reduces the number of required DFT data points.
Figure~\ref{fig:ndata} plots the convergence of the first-nearest-neighbor Fe--Fe exchange coupling $J_{01}$ and the magnitude of the DMI vector $|\vct D_{01}|$ versus the number of DFT data points for the torque-fitting and energy-fitting approaches.
In the conventional energy-fitting regression, both parameters begin to stabilize around $N_{\rm data}=\!196$, which is the number of independent parameters in the SCE model.
In contrast, the torque-fitting regression requires far fewer DFT data points for parameter convergence, because each configuration supplies $3N_{\rm a}$ torque components (the $x$, $y$, and $z$ components on each atom) rather than a single scalar energy.
A simple counting argument therefore suggests a potential reduction in the required number of configurations by a factor on the order of $3N_{\rm a}$ relative to energy fitting (i.e., $\sim 1/192$ for the present system with $N_{\rm a}=64$), although the practical gain depends on correlations among torque components, symmetry constraints, and the resulting conditioning of the design matrix.
Consistently, we observe that convergence is achieved with only a small number of data points.
Collectively, these reductions make the SCE model practical and data-efficient when parameterized via torque-fitting regression.
As the supercell size grows (and thus $N_{\rm a}$ increases), the relative efficiency over the energy-fitting approach becomes even more pronounced, making torque-fitting parameterization especially advantageous for larger supercells.

We comment on a small discrepancy, approximately 0.6~meV, between the converged values obtained by energy fitting and torque fitting in Fe--Fe $J_{01}$ [Fig.~\ref{fig:ndata}(a)].
A possible origin of the discrepancy is the difference in how contributions from the interstitial region enter the spin-model parameters.
In the present torque fitting, we define the local magnetic moments in terms of atomic spheres containing most of the magnetization density and use the torques acting on them in the fitting.
In the energy fitting, the local magnetic moments are defined in the same way, but the energy contributions from the interstitial region outside those atomic spheres are also reflected in the resulting spin-model parameters.
Thus, the difference is not so much a matter of one approach being superior to the other, but rather a consequence of a difference in viewpoint regarding how local magnetic moments and the torques acting on them are defined.

\subsection{Evaluation of micromagnetic parameters}
\label{ssec:micromagnetic}

\begin{figure*}
\centering
\includegraphics[width=130mm]{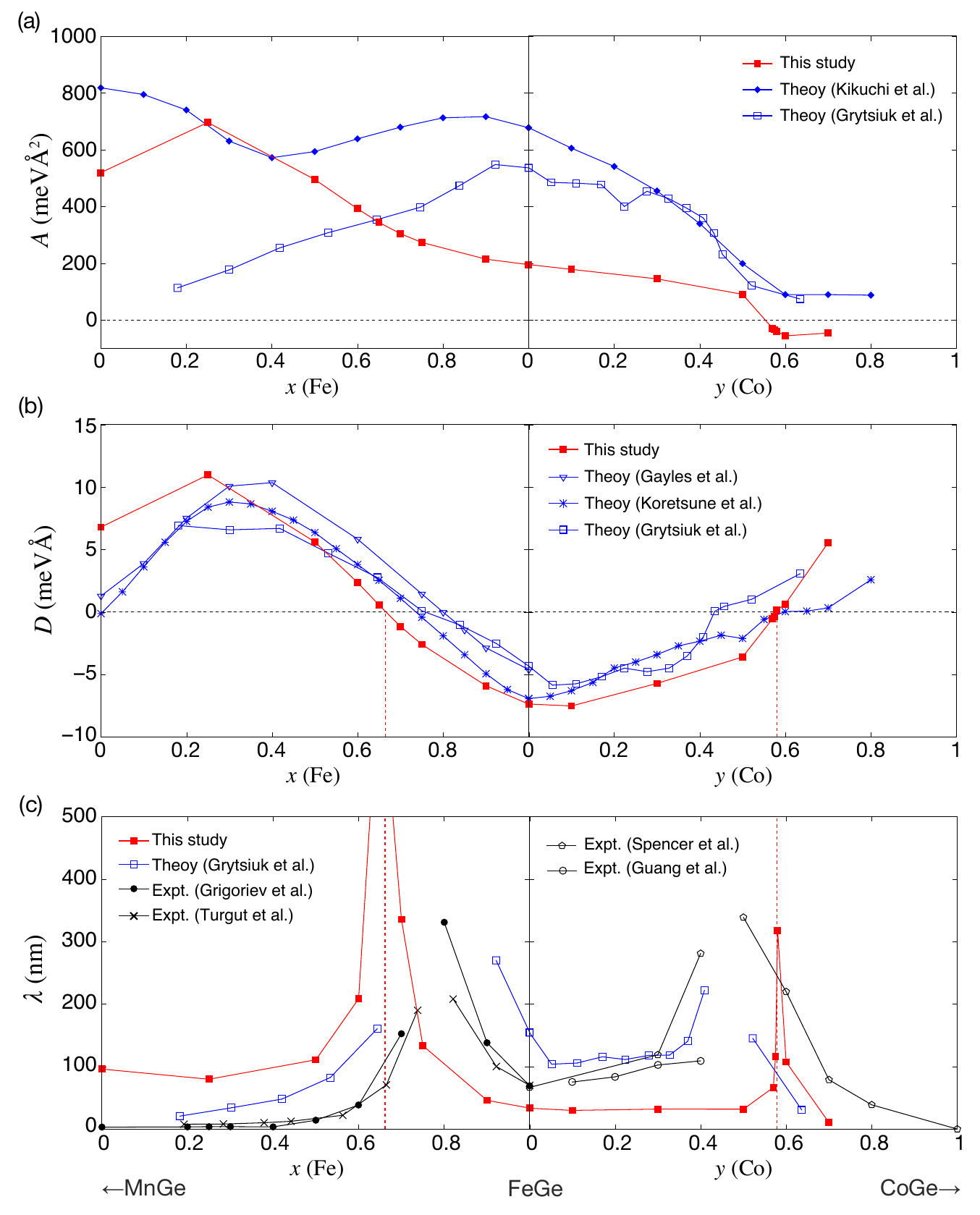}
\caption{(a) Spin stiffness constant $A$, (b) spiralization constant $D$, and (c) helical spin period $\lambda$ in ${\rm Mn}_{1-x}{\rm Fe}_{x}{\rm Ge}$ and ${\rm Fe}_{1-y}{\rm Co}_{y}{\rm Ge}$.
The two red dashed lines indicate the compositions at which $\lambda$ diverges in this study.
Results from previous theoretical works by Gayles et al.~\cite{Gayles2015-ow}, Kikuchi et al.~\cite{Kikuchi2016-iu}, Grytsiuk et al.~\cite{Grytsiuk2019-pa}, and Koretsune et al.~\cite{Koretsune2018-jh} as well as experimental works by Grigoriev et al.~\cite{Grigoriev2013-vb}, Turgut et al.~\cite{Turgut2018-ep}, Spencer et al.~\cite{Spencer2018-mh}, and Guang et al.~\cite{Guang2024-nc}, are also presented for comparison.}
\label{fig:ADL}
\end{figure*}

Figures~\ref{fig:ADL}(a)--(c) show the composition dependence of $A$, $D$, and $\lambda$, respectively.
The number of data $N_{\rm data}$ is set to 100.
For the spin stiffness constant $A$ [Fig.~\ref{fig:ADL}(a)], we find sizable deviations from previous theoretical studies~\cite{Kikuchi2016-iu,Grytsiuk2019-pa} across the composition range of ${\rm Mn}_{1-x}{\rm Fe}_{x}{\rm Ge}$.
Given that the experimental magnetic transition temperatures of $\rm FeGe$ and $ \rm MnGe$ are $\approx 280$~K~\cite{Lebech1989-eg} and $\approx 170$~K \cite{Kanazawa2011-ks}, respectively, and that within mean-field theory critical temperature is proportional to $A$, the results of Grytsiuk \textit{et al.}~\cite{Grytsiuk2019-pa} capture the expected trend (they evaluated spin-model parameters by a perturbative approach), whereas our data exhibit the opposite tendency.
A likely origin is the limited simulation cell used here (the 2$\times$2$\times$2 supercell).
Indeed, Ref.~\cite{Mendive-Tapia2021-yg} reports that the nearest-neighbor $J_{ij}$ between Mn atoms in MnGe is roughly twice as large as the nearest-neighbor Fe--Fe $J_{ij}$ in $\rm FeGe$; nevertheless, the smaller transition temperature of MnGe can be understood as arising from contributions of longer-range $J_{ij}$.
However, within the SCE framework this discrepancy can be systematically reduced by increasing the supercell size; the remaining choice becomes a trade-off between computational cost and accuracy.

Regarding the parameter $D$ [Fig.~\ref{fig:ADL}(b)], its composition dependence agrees well with previous theoretical studies.
In particular, we find the sign change of $D$ around $x(\mathrm{Fe}) \approx 0.6$--$0.7$ and $y(\mathrm{Co}) \approx 0.6$.
As noted above, the helical period $\lambda$ diverges at compositions where $D=0$.
Figure~\ref{fig:ADL}(c) compares $\lambda$ calculated from $A$ and $D$ as Eq. (\ref{eq:lambda}) with experimental results. Experiments likewise show a divergence near $x(\mathrm{Fe}) \approx 0.8$ and $y(\mathrm{Co}) \approx 0.5$; aside from modest offsets, the overall trends are consistent.
These offsets are reasonable in light of our computational setup---varying the electron number only while fixing the structural parameters---and fall within the expected uncertainty.
In Supplementary Materials Sec.~S2, we discuss the convergence of $A$, $D$, and $\lambda$ for FeGe and $(\mathrm{Mn}_{0.35}\mathrm{Fe}_{0.65})$Ge, the latter corresponding to the composition at which $\lambda$ diverges~\cite{SI}.
We also present in Supplementary Materials Sec.~S3 the composition dependences of $\lambda$ and the magnetic transition temperature obtained within the mean-field approximation following Ref.~\cite{Mendive-Tapia2021-yg}, which confirms that the spin-spiral state is stabilized also at the mean-field level~\cite{SI}.
Overall, although finite-size effects associated with the supercell used here are present, our calculations indicate that the SCE framework attains systematically improved accuracy as the cell size is increased.
In the following, we evaluate the impact of spin fluctuations---a representative temperature effect---where the advantages of SCE become particularly pronounced.

\subsection{Higher-order interactions for thermally disordered states}
\label{higher-order}
Whether a spin model can accurately capture the impact of thermal magnetic disorder is a key requirement for quantitative modeling.
Following the $\tau=0.1$ results presented above, we increase the reduced temperature to $\tau=0.3$, sample 100 spin configurations, and construct the corresponding SCE model.
We also construct an SCE model with the expansion cutoff in Eq.~(\ref{eq:SCE_fege}) raised from $l_{\max}=1$ to $l_{\max}=2$ as
\begin{align}
\begin{split}
E_{\rm SCE} =
J_0 + \sum_{i, j}\sum_{l_i, l_j = 1}^{l_{\rm max} = 2} \sum_{m_i= -l_i}^{l_i} \sum_{m_j= -l_j}^{l_j} J_{ij\vct{l}\vct{m}} Z_{l_i m_i}(\hat{\vct{e}}_i) Z_{l_j m_j}(\hat{\vct{e}}_j),
\end{split}
\label{eq:SCE_fege_lmax2}
\end{align}
which, in conventional spin-model language, is equivalent to 
\begin{align}
	\begin{split}
	E = &-\sum_{i, j} J_{ij} \hat{\vct{e}}_i \cdot \hat{\vct e}_j \\
	&+ \sum_{i, j} {\vct D}_{ij} \cdot(\hat{\vct{e}}_i \times \hat{\vct e}_j)\\
	&+ \sum_{i, j} {\hat{\vct{e}}_i}^{\mathsf{T}} {\vct \Gamma}_{ij} \hat{\vct e}_j\\
	&+ \sum_{i, j} B_{ij} (\hat{\vct{e}}_i \cdot \hat{\vct{e}}_j)^2 \\
	&+ H^{(4)}_{\rm aniso},
	\end{split}
\end{align}
where $B_{ij}$ is the isotropic biquadratic interaction parameter and $H^{(4)}_{\rm aniso}$ collects SOC-induced quartic anisotropic two-spin interactions.

\begin{figure*}
\centering
\includegraphics[width=\linewidth]{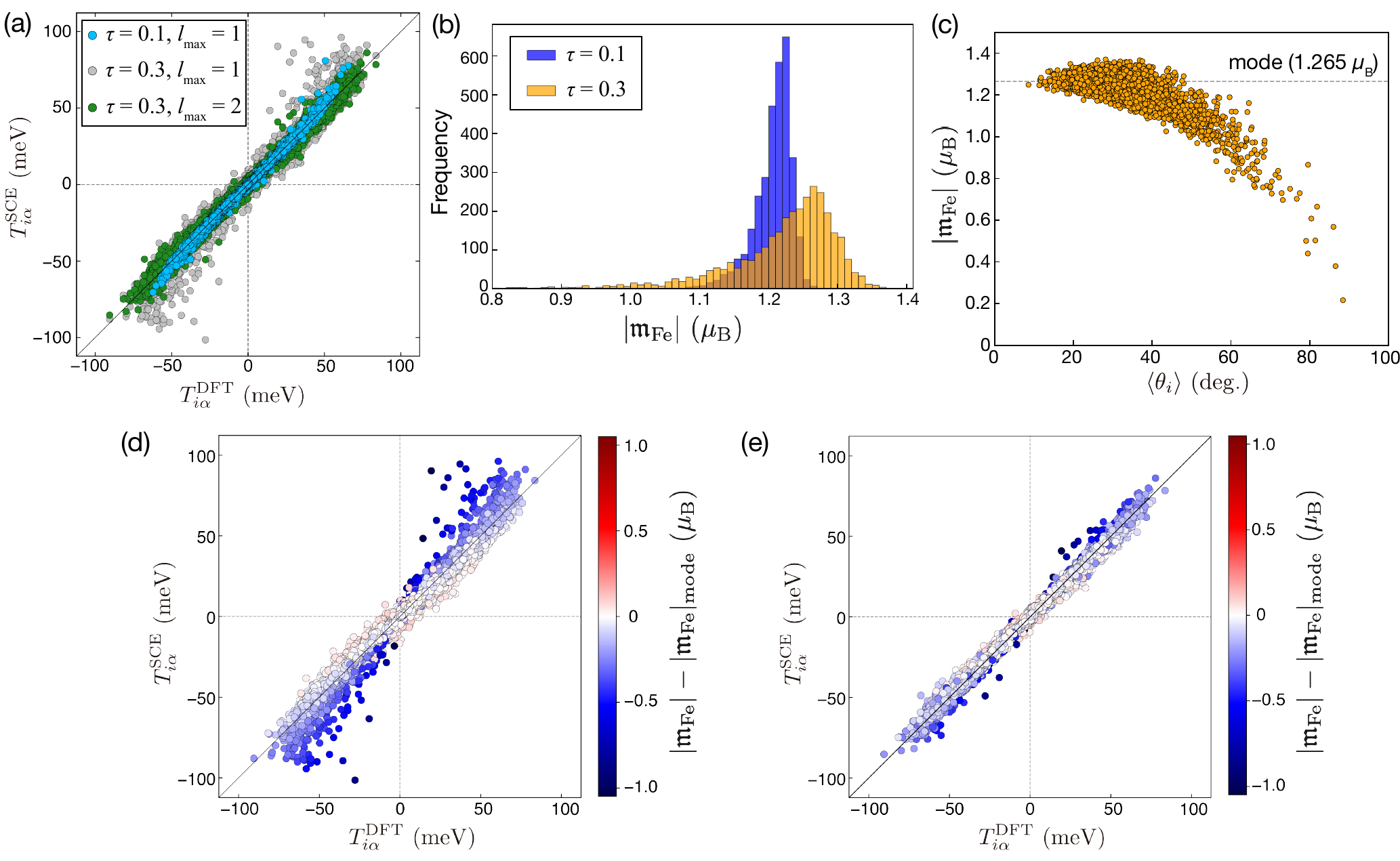}
\caption{Parity plot of site-resolved torque components: the horizontal axis shows $T_{{\rm DFT}, i\alpha}$ and the vertical axis shows $T_{{\rm SCE}, i\alpha}$, where $\alpha \in \{x,y,z\}$.
(b) Histogram of Fe-moment magnitude $|\mathfrak{m}_{\rm Fe}|$ at $\tau = 0.1$ (blue) and $\tau = 0.3$ (orange); bin width $0.01\ \mu_{\rm B}$.
(c) $|\mathfrak{m}_{\rm Fe}|$ as a function of $\langle \theta_i \rangle$ defined in Eq. (\ref{eq:relangle}).
(d, e) Parity plots of Fe torques at $\tau = 0.3$ for $l_{\rm max}=1$ and $l_{\rm max}=2$, respectively (same axis definitions as in panel (a)).
In (d) and (e), points are color-coded by the deviation from the mode of Fe moment distribution, $|\mathfrak{m}_{\rm Fe}| - |\mathfrak{m}_{\rm Fe}|_{\rm mode}$with $|\mathfrak{m}_{\rm Fe}|_{\rm mode}=1.265\ \mu_{\rm B}$.}
\label{fig:temp_vary}
\end{figure*}

The blue and gray points in Figure~\ref{fig:temp_vary}(a) compare DFT torques with those from the $l_{\max}=1$ SCE model at $\tau=0.1$ and $\tau=0.3$, respectively, with all other settings identical, thereby isolating the effect of increased magnetic disorder.
At $\tau=0.1$, the $l_{\max}=1$ SCE reproduces the DFT torques closely, whereas its predictive accuracy becomes less accurate at $\tau=0.3$.
These trends indicate that truncation at $l_{\max}=1$ is valid at very low temperatures but insufficient at elevated temperatures.
By contrast, at $\tau=0.3$, extending the SCE to $l_{\max}=2$ substantially improves the torque prediction accuracy relative to $l_{\max}=1$.
We attribute the reduced $l_{\max}=1$ accuracy at $\tau=0.3$ to the larger spread in the magnetic moment magnitude of Fe, $|\mathfrak{m}_{\rm Fe}|$, as discussed below.

Figure~\ref{fig:temp_vary}(b) shows histograms of the Fe magnetic-moment magnitude, $|\mathfrak{m}_{\rm{Fe}}|$, at $\tau=0.1$ and $\tau=0.3$ (the Ge moments are negligible compared with those of Fe and are therefore omitted).
The $\tau=0.3$ distribution is visibly broader than at $\tau=0.1$, indicating larger dispersion in $|\mathfrak{m}_{\rm{Fe}}|$.
This broad variance in $|\mathfrak{m}_{\rm{Fe}}|$ is strongly correlated with local magnetic disordering.
To quantify short-range order among Fe--Fe pairs, we define for each Fe site $i$ the average relative angle to its first-nearest Fe neighbors,
\begin{align}
\langle \theta_i \rangle = \frac{1}{N_{\rm 1NN}(i)}\sum_{j\in {\rm 1NN}(i)} \arccos\left(\hat{\bm e}_i \cdot \hat{\bm e}_j\right),
\label{eq:relangle}
\end{align}
where ${\rm 1NN}(i)$ is the set of first-nearest Fe neighbors of site $i$ and $N_{\rm 1NN}(i)$ is the number of Fe atoms in the set ${\rm 1NN}(i)$.
Figure~\ref{fig:temp_vary}(c) clearly shows that larger $\langle \theta_i \rangle$ leads to the shrinkage of $|\mathfrak{m}_{\rm{Fe}}|$,
i.e., $|\mathfrak{m}_{\rm{Fe}}|$ strongly depends on local magnetic order.
Figure~\ref{fig:temp_vary}(d) presents the $\tau=0.3$, $l_{\max}=1$ parity plot with points color-coded by the deviation from the modal Fe moment magnitude at $\tau=0.3$, $|\mathfrak{m}_{\mathrm{Fe}}|-|\mathfrak{m}_{\mathrm{Fe}}|_{\mathrm{mode}}$.
Taken together, these results indicate that the degradation of the $l_{\max}=1$ SCE torque predictions with increasing $\tau$ stems from its inability to capture the temperature-dependent spread in local-moment magnitudes $|\mathfrak{m}_{\rm Fe}|$ induced by magnetic disordering.
Moreover, the improvement of torque predictive accuracy by including higher-order interactions can be understood intuitively from Fig.~\ref{fig:temp_vary}(c); these terms provide the nonlinear angular dependence needed to reproduce the energy and torque for configurations with large relative spin angles, which cannot be captured by the low-order $l_{\max}=1$ model.
Indeed, the torques on atoms with strongly shrunk local moments---poorly captured by the $l_{\max}=1$ model [Fig.~\ref{fig:temp_vary}(d)]---are reproduced much more accurately by the $l_{\max}=2$ SCE model [Fig.~\ref{fig:temp_vary}(e)].

Experimentally, while the helical period $\lambda$ of $\rm FeGe$ exhibits only weak temperature dependence from a quite low temperature to the critical temperature, the spin-stiffness constant $A$ decreases with increasing temperature~\cite{Siegfried2017-qq}.
Clearly, this implies that an exchange-DMI-only model, with parameters determined at a single temperature, is insufficient to quantitatively describe magnetic order over a broad temperature range.
To address this issue, two approaches can be considered: (i) employ a temperature-dependent exchange-DMI-only model or (ii) construct a spin model that includes higher-order interactions capable of describing a broader temperature range (i.e., a broader spin configuration space).
However, as we have shown, the $l_{\max}=1$ SCE model cannot adequately renormalize the effects of fluctuations in the local-moment magnitude $|\mathfrak{m}_{\rm Fe}|$; thus, option (ii) is the more appropriate route for capturing magnetism across a wide temperature range.
In this context, the advantages of SCE become particularly pronounced: by tuning the model capacity---e.g., increasing $l_{\max}$ and/or the set of included clusters---one can systematically build quantitative spin models.
In practice, care must be taken to monitor and mitigate overfitting during model selection.

Finally, we note a conceptual distinction from approaches based on the magnetic force theorem~\cite{Liechtenstein1987-ev,Oguchi1983-vi}.
Whereas MFT extracts interaction parameters from the change in single-particle energy induced by infinitesimal spin rotations about a chosen reference state within a frozen potential, the SCE method samples finite spin rotations and fits quantities obtained from fully self-consistent DFT calculations, thereby reflecting the total DFT energy.
Therefore, provided that finite-size effects are adequately controlled, SCE offers a complementary self-consistent route to estimating higher-order couplings, especially in regimes where finite rotations or longitudinal fluctuations are relevant.
 

\section{Conclusion}
\label{sec:discussion}
We have proposed a supercell-based framework for evaluating magnetic interactions of arbitrary form within SCE.
In this approach, diverse spin configurations and their site-resolved magnetic torques are computed by constrained noncollinear-spin DFT and used as training data, enabling a highly efficient torque-fitting regression for the SCE model.
We applied this framework to representative chiral magnet systems, B20-type $\rm FeGe$ and its variants ${\rm Mn}_{1-x}{\rm Fe}_{x}{\rm Ge}$ and ${\rm Fe}_{1-y}{\rm Co}_{y}{\rm Ge}$.
The computed spiralization constant $D$ agrees with other theoretical studies, and the compositions where the helical period diverges are also correctly reproduced.
As a distinctive strength of SCE, we have demonstrated the importance of higher-order interactions in moderately disordered magnetic states of $\rm FeGe$.
At elevated temperature, the Fe magnetic moments exhibit a wide variance around the modal value.
As a result, the lowest-order truncation fails to maintain torque predictive accuracy, whereas including higher-order terms restores agreement.
The variance of the local moment strongly depends on the local magnetic disorder (i.e., relative spin angles between nearest-neighbor spins), and higher-order interactions play a compensating role when these relative angles become large.
Thus, although the SCE model does not include an explicit degree of freedom for the moment magnitude, spin-configuration-dependent changes in the moment length are implicitly captured through higher-order interactions.

A variety of extensions of the SCE framework are conceivable.
In this work we preselected a magnetic state of interest and sampled spin configurations in its vicinity; however, because the SCE basis is complete and orthogonal, it can---within the regime where the mapping from spin configuration to energy is single-valued---represent the energies of arbitrary magnetic states.
When combined with strategies that efficiently sample a broad manifold of configurations, this should enable quantitative descriptions over wide temperature ranges and high-fidelity predictions of more complex magnetic textures.
While the use of supercells increases computational cost, it also enables direct treatments of atomic displacements.
Since spin and lattice degrees of freedom can couple strongly (e.g., via lattice vibrations~\cite{Mankovsky2020-ej, Ruban2018-mb, Mauger2014-lq, Tanaka2020-vf, Tanaka2020-qe} and magneto-elastic effects~\cite{Lu2015-oy, Mankovsky2023-er, Miranda2025-ru}), a supercell approach that explicitly displaces atoms offers a route to computing spin--lattice coupling with higher accuracy.

\section*{Appendix}
\subsection*{Conversion from the SCE model to a conventional spin model}
\label{ssec:conversion}
For the $l=1$ sector we relabel the real spherical harmonics by Cartesian indices via
$Z_{1x}\equiv Z_{11}$, $Z_{1y}\equiv Z_{1\bar1}$, and $Z_{1z}\equiv Z_{10}$, and define
$\cvec{Z}_1(\hat{\vct e}_i) \equiv (Z_{1x}(\hat{\vct e}_i), Z_{1y}(\hat{\vct e}_i), Z_{1z}(\hat{\vct e}_i))^{\mathsf T}$.
Using
\[
\sqrt{4\pi/3}\ \cvec{Z}_1(\hat{\vct e}_i)=\hat{\vct e}_i,
\]
the two-body $l=1$ contribution of the SCE energy can be written as
\begin{align}
E^{(2)}_{l=1}
&= 4\pi \sum_{i, j} \cvec{Z}_1(\hat{\vct e}_i)^{\mathsf T}\, \mat{J}^{(1)}_{ij}\, \cvec{Z}_1(\hat{\vct e}_j) \\
&= 3\sum_{i, j}\hat{\vct e}_i^{\mathsf T}\,\mat{J}^{(1)}_{ij}\,\hat{\vct e}_j
\;\equiv\; \sum_{i, j}\hat{\vct e}_i^{\mathsf T}\,\mathcal{J}_{ij}\,\hat{\vct e}_j,
\end{align}
with the identification
\begin{align}
\mathcal{J}_{ij}=3\,\mat{J}^{(1)}_{ij},\qquad
\mat{J}^{(1)}_{ij}=\bigl(J^{(1)}_{ij,\mu\nu}\bigr)_{\mu,\nu\in\{x,y,z\}}.
\end{align}
Thus the conventional exchange tensor $\mathcal{J}_{ij}$ is obtained directly from the $l=1$ block of the SCE coefficients.  Decomposing $\mathcal{J}_{ij}$ into isotropic, symmetric traceless, and antisymmetric parts gives
\begin{align}
J_{ij} &= -\frac{1}{3}\,\mathrm{tr}\,\mathcal{J}_{ij},\\
{\vct \Gamma}_{ij} &= \frac{1}{2}\bigl(\mathcal{J}_{ij}+\mathcal{J}_{ij}^{\mathsf T}\bigr)+J_{ij}\,\mathbb{I},\\
\mathcal{W}_{ij} &= \frac{1}{2}\bigl(\mathcal{J}_{ij}-\mathcal{J}_{ij}^{\mathsf T}\bigr),
\end{align}
where $\mathbb{I}$ is the identity matrix.
The antisymmetric exchange $\mathcal{W}_{ij}$ is in one-to-one correspondence with the DMI vector $\vct D_{ij}$.
At the component level,
\begin{align}
\mathcal W_{ij}^{\alpha\beta} &= \sum_{\gamma}\varepsilon_{\alpha\beta\gamma}\, D_{ij}^{\gamma},
\label{eq:W_from_D}\\
D_{ij}^{\gamma} &= \frac{1}{2}\sum_{\alpha\beta}\varepsilon_{\alpha\beta\gamma}\,\mathcal W_{ij}^{\alpha\beta},
\label{eq:D_from_W}
\end{align}
where $\varepsilon_{\alpha\beta\gamma}$ is the Levi--Civita symbol.
Combining with $\mathcal{W}_{ij}=\tfrac12(\mathcal{J}_{ij}-\mathcal{J}_{ij}^{\mathsf T})$, we obtain
\begin{align}
D_{ij}^{\gamma}
= \frac{1}{4}\sum_{\alpha\beta}\varepsilon_{\alpha\beta\gamma}\,
\bigl(\mathcal{J}_{ij}^{\alpha\beta}-\mathcal{J}_{ij}^{\beta\alpha}\bigr).
\end{align}
In matrix form, Eq.~\eqref{eq:W_from_D} is equivalent to
\begin{align}
\mathcal W_{ij} =
\begin{pmatrix}
0 & D_{ij}^{z} & -\,D_{ij}^{y}\\
-\,D_{ij}^{z} & 0 & D_{ij}^{x}\\
D_{ij}^{y} & -\,D_{ij}^{x} & 0
\end{pmatrix}.
\label{eq:W_matrix}
\end{align}
Moreover, for any vectors $\vct a,\vct b$ one has the identity
\begin{align}
\vct a^{\mathsf T}\,\mathcal W_{ij}\,\vct b \;=\; \vct D_{ij}\cdot(\vct a\times\vct b),
\label{eq:bilinear_identity}
\end{align}
which makes the equivalence of the antisymmetric exchange term
$\hat{\vct e}_i^{\mathsf T}\mathcal W_{ij}\hat{\vct e}_j$
and the DMI form $\vct D_{ij}\!\cdot\!(\hat{\vct e}_i\times\hat{\vct e}_j)$ explicit.

\section*{Data availability}
The data supporting the findings of this study are available from the first author, T.T., upon reasonable request.

\section*{Code availability}
The methodology and implementation presented in this study are available as an open-source Julia package, \textit{Magesty.jl}, on the first author's GitHub: \url{https://github.com/Tomonori-Tanaka/Magesty.jl}.

\begin{acknowledgments}
This work was partly supported by JSPS KAKENHI Grant Number JP24K01144 and MEXT-DXMag Grant Number JPMXP1122715503. The calculations were partly carried out by using facilities of the Supercomputer Center at the Institute for Solid State Physics, the University of Tokyo, and TSUBAME4.0 supercomputer at Institute of Science Tokyo.
\end{acknowledgments}



\end{document}


\title{Supplemental Material:\\General spin models from noncollinear spin density functional theory and spin-cluster expansion}
\author{Tomonori Tanaka}
 \email{tanaka.t.da74@m.isct.ac.jp}
\author{Yoshihiro Gohda}
 \email{gohda@mct.isct.ac.jp}
\affiliation{Department of Materials Science and Engineering, Institute of Science Tokyo, Yokohama 226-8501, Japan}
\date{\today}
\maketitle
\tableofcontents
\renewcommand{\thesection}{S\arabic{section}}
\section{Real spherical harmonics in Cartesian coordinates}
Here, we summarize the derivation of the Cartesian forms of the real spherical harmonics and their derivatives.
The real spherical harmonics $Z_{lm}$ are defined in terms of the complex $Y_{lm}$ as
\begin{equation}
Z_{lm} =
\begin{dcases}
\cfrac{(-1)^m}{\sqrt{2}}\!\left(Y_{lm} + (-1)^{m} Y_{l\bar m}\right), & m > 0,\\[4pt]
Y_{l0}, & m = 0,\\[4pt]
\cfrac{(-1)^{m}}{i\sqrt{2}}\!\left(Y_{l|m|} - (-1)^{|m|} Y_{l\overline{|m|}}\right), & m < 0,
\end{dcases}
\label{transform}
\end{equation}
and the Cartesian form of the complex spherical harmonics $Y_{lm}$ for $m \ge 0$ is \cite{Drautz2020-ao}
\begin{align}
	Y_{lm}(\hat{{\bm e}}_i) &= (e_{ix} + ie_{iy})^m\bar{P}_{lm}(e_{iz}),\label{eq:y_cart}\\
	\bar{P}_{lm}(e_{iz}) &= (-1)^m\sqrt{\cfrac{2l+1}{4\pi}\cdot\cfrac{(l-m)!}{(l+m)!}}\cfrac{\dif^m}{\dif e_{iz}^m}P_{l}(e_{iz}),
\end{align}
where $\bar m \equiv -m$ and $\overline{|m|} \equiv -|m|$.
For $m<0$, the corresponding expressions follow from the relation $Y_{l\bar m} = (-1)^{|m|} Y_{l|m|}^{*}$.

For $m>0$, suppressing the site index and writing $(e_x,e_y,e_z)$ for the Cartesian components,
\begin{align}
\begin{split}
Z_{lm} &= \cfrac{(-1)^m}{\sqrt{2}}\left(Y_{lm} + (-1)^{m}Y_{l\bar{m}}\right) \\
&=  \cfrac{(-1)^m}{\sqrt{2}}\left(Y_{lm} + {Y_{lm}}^{*}\right) \\
&= \cfrac{(-1)^m}{\sqrt{2}}\bar{P}_{lm}(e_{z})\left\{(e_{x} + ie_{y})^{m} + (e_{x} - ie_{y})^{m}\right\}\\
&= \cfrac{(-1)^m}{\sqrt{2}}\bar{P}_{lm}(e_{z}) \sum^{m}_{k=0} \binom{m}{k} e_{x}^{m-k} e_{y}^{k} (i^k + (-i)^k) \\
&= (-1)^{m}\sqrt{2} \bar{P}_{lm}(e_{z}) \sum^{\left[ m/2 \right]}_{k=0} (-1)^{k} \binom{m}{2k}e_{x}^{m-2k}e_{y}^{2k}, 
\end{split}
\end{align}
where $[x]$ denotes the floor function (i.e., the largest integer less than or equal to $x$), and $\binom{m}{k}$ is the binomial coefficient.
Similarly, for $m<0$, 
\begin{align}
  \begin{split}
Z_{lm} &= \cfrac{(-1)^{m}}{i\sqrt{2}}\left(Y_{l|m|} - (-1)^{|m|}Y_{l\overline{|m|}}\right) \\
&= \cfrac{(-1)^{m}}{i\sqrt{2}} \left(Y_{l|m|} - Y^{*}_{l|m|}\right) \\
&= (-1)^{m}\sqrt{2} \bar{P}_{l|m|}(e_{z}) \sum^{\left[(|m|-1)/2 \right]}_{k=0} (-1)^{k} \binom{|m|}{2k+1}e_{x}^{|m|-(2k+1)}e_{y}^{2k+1}.
  \end{split}
\end{align}
For $m=0$,
\begin{align}
	Z_{l0} &= Y_{l0} = \bar{P}_{l0}(e_z).
\end{align}
In the above derivations we used the identities
\begin{align}
i^k + (-i)^k &= 
\begin{dcases}
0 & \text{if } k \text{ is odd} \\
2(-1)^{k/2} & \text{if } k \text{ is even}
\end{dcases},\\
i^k - (-i)^k &= 
\begin{dcases}
2i(-1)^{(k+1)/2} & \text{if } k \text{ is odd} \\
0 & \text{if } k \text{ is even}
\end{dcases}.
\end{align}

The Cartesian derivatives then follow directly.
For $m > 0$,
\begin{align}
  \begin{split}
\frac{\partial Z_{lm}}{\partial e_x} &= \frac{(-1)^m}{\sqrt{2}}\left( \frac{\partial Y_{lm}}{\partial e_x} + \frac{\partial Y^{*}_{lm}}{\partial e_x} \right) \\
&= \frac{(-1)^m}{\sqrt{2}} \bar{P}_{lm}(e_{z}) \left[ m(e_{x} + ie_{y})^{m-1} + m(e_{x} - ie_{y})^{m-1} \right] \\
&= \frac{(-1)^m}{\sqrt{2}} m\bar{P}_{lm}(e_{z}) \sum^{m-1}_{k=0} \binom{m-1}{k} e_{x}^{m-1-k} e_{y}^{k} (i^k + (-i)^k) \\
&= (-1)^m \sqrt{2} m\bar{P}_{lm}(e_{z}) \sum^{[(m-1)/2]}_{k=0}(-1)^k \binom{m-1}{2k} e_{x}^{m-1-2k} e_{y}^{2k},\\
  \end{split}
\end{align}
\begin{align}
  \begin{split}
\frac{\partial Z_{lm}}{\partial e_y} &= \frac{(-1)^m}{\sqrt{2}}\left( \frac{\partial Y_{lm}}{\partial e_y} + \frac{\partial Y^{*}_{lm}}{\partial e_y} \right) \\
&= \frac{(-1)^m}{\sqrt{2}} \bar{P}_{lm}(e_{z}) \left[ im(e_{x} + ie_{y})^{m-1} - im(e_{x} - ie_{y})^{m-1} \right] \\
&= \frac{(-1)^m}{\sqrt{2}} im\bar{P}_{lm}(e_{z}) \sum^{m-1}_{k=0} \binom{m-1}{k} e_{x}^{m-1-k} e_{y}^{k} (i^k - (-i)^k)\\
&= (-1)^{m+1} \sqrt{2} m \bar{P}_{lm}(e_{z}) \sum^{[(m-2)/2]}_{k=0} (-1)^k \binom{m-1}{2k+1} e_{x}^{m-2-2k} e_{y}^{2k+1},\\
  \end{split}
\end{align}
\begin{align}
  \begin{split}
\frac{\partial Z_{lm}}{\partial e_z} &= (-1)^m \sqrt{2} \left( \cfrac{\dif \bar{P}_{l m}(e_z)}{\dif e_z} \right)
\sum_{k=0}^{[m/2]} (-1)^k \binom{m}{2k} e_x^{m - 2k} e_y^{2k}.
  \end{split}
\end{align}
For $m = 0$,
\begin{align}
	\frac{\partial Z_{l0}}{\partial e_x} &=  0,\\
	\frac{\partial Z_{l0}}{\partial e_y} &=  0,\\
	\frac{\partial Z_{l0}}{\partial e_z} &=  \frac{\dif \bar{P}_{l0}(e_z)}{\dif e_z}.
\end{align}
For $m < 0$,
\begin{align}
  \begin{split}
\frac{\partial Z_{lm}}{\partial e_x} &= \frac{(-1)^m}{i\sqrt{2}}\left( \frac{\partial Y_{l|m|}}{\partial e_x } - \frac{\partial Y_{l\overline{|m|}}}{\partial e_x } \right) \\
&= \frac{(-1)^m}{i\sqrt{2}} \bar{P}_{l|m|}(e_{z}) \left[ |m|(e_{x} + ie_{y})^{|m|-1} - |m|(e_{x} - ie_{y})^{|m|-1} \right]\\
&= \frac{(-1)^m}{i\sqrt{2}} |m| \bar{P}_{l|m|}(e_{z}) \sum^{|m|-1}_{k=0} \binom{|m|-1}{k} e_{x}^{|m|-1-k} e_{y}^{k} (i^k - (-i)^k) \\
&= (-1)^m \sqrt{2} |m| \bar{P}_{l|m|}(e_{z}) \sum^{[(|m|-2)/2]}_{k=0} (-1)^k \binom{|m|-1}{2k+1} e_{x}^{|m|-2-2k} e_{y}^{2k+1},
\end{split}
\end{align}
\begin{align}
  \begin{split}
\frac{\partial Z_{lm}}{\partial e_y} &= \frac{(-1)^m}{i\sqrt{2}}\left( \frac{\partial Y_{l|m|}}{\partial e_y } - \frac{\partial Y_{l\overline{|m|}}}{\partial e_y } \right)\\
&= \frac{(-1)^m}{\sqrt{2}}|m| \bar{P}_{l|m|}(e_{z}) \sum^{|m|-1}_{k=0} \binom{|m|-1}{k} e_{x}^{|m|-1-k} e_{y}^{k} (i^k + (-i)^k)\\
&= (-1)^m \sqrt{2} |m| \bar{P}_{l|m|}(e_{z}) \sum^{[(|m|-1)/2]}_{k=0} (-1)^k \binom{|m|-1}{2k} e_{x}^{|m|-1-2k} e_{y}^{2k},\\
  \end{split}
\end{align}
\begin{align}
\frac{\partial Z_{lm}}{\partial e_z} &= (-1)^m \sqrt{2} \left( \cfrac{\dif \bar{P}_{l |m|}(e_z)}{\dif e_z} \right) \sum_{k=0}^{[(|m| - 1)/2]} (-1)^k \binom{|m|}{2k + 1} e_x^{|m| - 2k - 1} e_y^{2k + 1}.
\end{align}

\clearpage
\section{Convergence of micromagnetic parameters}
\begin{figure}[h]
	\includegraphics[width=120mm]{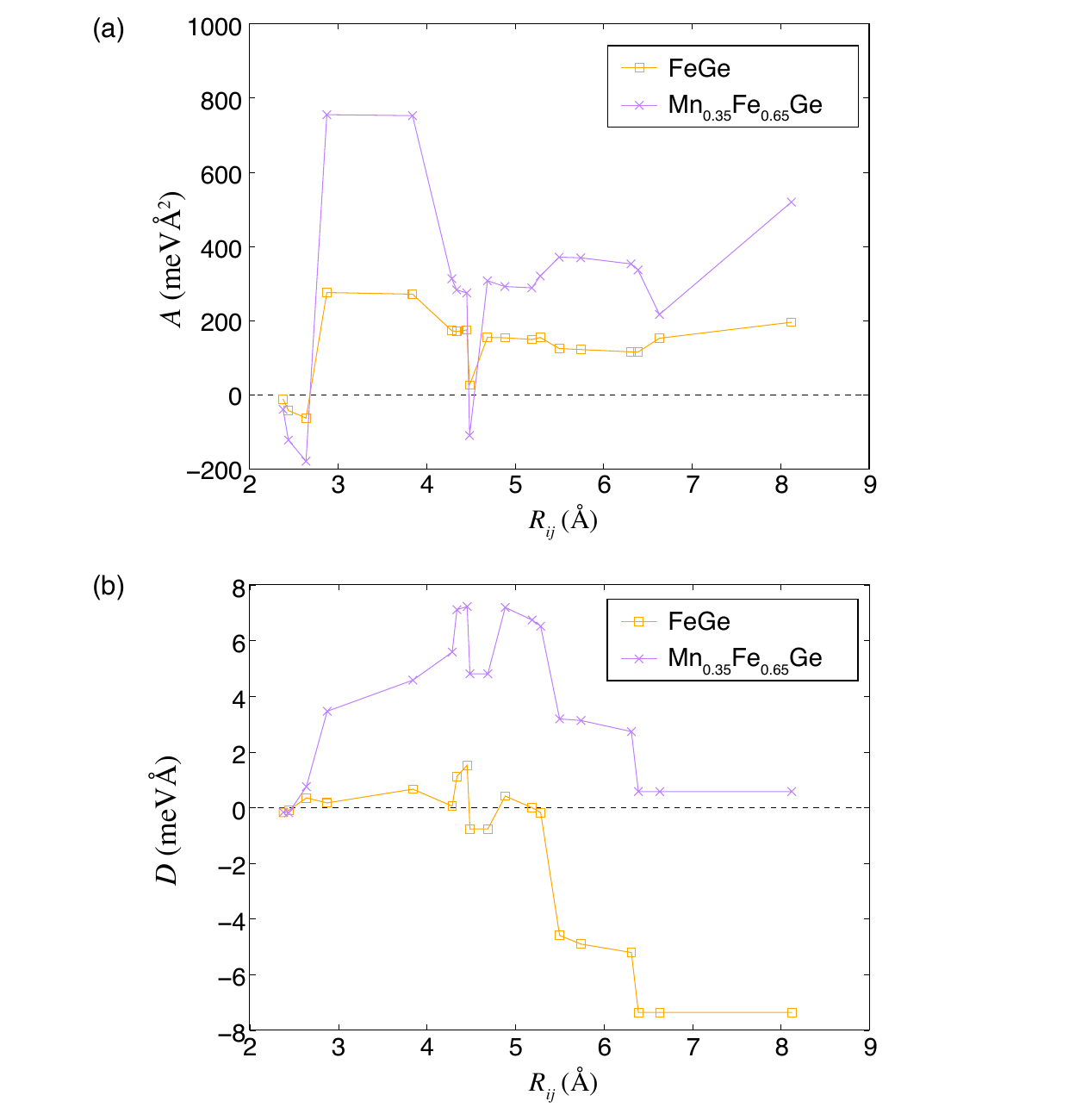}
	\caption{Convergence of the micromagnetic parameters (a) $A$ and (b) $D$ with respect to the cutoff distance for the magnetic interactions in FeGe and Mn$_{0.35}$Fe$_{0.65}$Ge.}
\label{fig:convergence}
\end{figure}

We show the convergence of the micromagnetic parameters with respect to the cutoff of the magnetic interactions in $\mathrm{FeGe}$ and $\mathrm{Mn}_{0.35}\mathrm{Fe}_{0.65}\mathrm{Ge}$, the latter corresponding to the composition at which the helical spin period $\lambda$ diverges.
The spin-stiffness constant $A$ and the spiralization constant $D$ are defined as 
\begin{align}
A &= \cfrac{1}{2}\sum_{i, j} J_{ij} {\vct R}_{ij} \cdot {\vct R}_{ij},\\
D &= \sum_{i, j} {\vct D}_{ij} \cdot{\vct R}_{ij}.
\end{align}
Here, we vary the cutoff of the interactions included in the summation, and plot the resulting $A$ and $D$ in Fig. \ref{fig:convergence}.
For $A$, we find that the convergence is relatively poor for $\mathrm{Mn}_{0.35}\mathrm{Fe}_{0.65}\mathrm{Ge}$ because of the contribution from long-range $J_{ij}$, whereas it is reasonably good for $\mathrm{FeGe}$.
This is likely related to the point discussed in the main text: Ref.~\cite{Mendive-Tapia2021-yg} reported that, in MnGe, long-range pairwise $J_{ij}$ interactions make a significant contribution to $A$.
Therefore, the present result suggests that highly accurate calculations for Mn-containing systems require larger supercells, which remains an important subject for future work.

\clearpage
\section{Mean-field analysis}
In the main text, we evaluated the helical spin period $\lambda$ via the mapping to the micromagnetic model.
Here, we perform a mean-field analysis to evaluate the helical spin period $\lambda$, as an alternative approach.
The analysis method employed here is based on Ref.~\cite{Mendive-Tapia2021-yg}; in that work, the magnetic free energies of MnGe and FeGe were analyzed using a spin model derived from the Curie-Weiss paramagnetic state.
In contrast, here we use spin-model parameters derived from sampling near the ferromagnetic state ($\tau = 0.1$).
Below, we first summarize the mean-field analysis method following Ref.~\cite{Mendive-Tapia2021-yg}.

The classical spin model considered here is given by
\begin{align}
  E
  =
  -\sum_{\RR,\RR'}\sum_{\mu\nu}
  \ee_{\RR\mu}^{\mathsf T}
  \mathcal{J}_{\mu\nu}(\RR'-\RR)
  \ee_{\RR'\nu},
  \label{eq:H_real}
\end{align}
where $\mathcal{J}$ represents the exchange coupling tensor defined in the Appendix of the main text, $\RR$ is a Bravais lattice cell vector, and $\mu, \nu (= 1, ..., n_{\rm sub})$ are sublattice indices.
The exchange coupling tensor $\mathcal{J}$ satisfies the following symmetry relations:
\begin{align}
\mathcal{J}_{\mu\nu}(-\bm \Delta) = \mathcal{J}_{\nu\mu}^{\mathsf T}(\bm \Delta),
\qquad \bm\Delta=\RR'-\RR .
\label{eq:reality}
\end{align}

We introduce the local order parameter
\begin{align}
{\bf m}_{\RR\mu} = \langle \ee_{\RR\mu} \rangle,
\end{align}
and decompose the spin direction into its mean value and fluctuation,
\begin{align}
\ee_{\RR\mu} = {\bf m}_{\RR\mu} + \delta \ee_{\RR\mu}.
\end{align}
For one interaction term, with $\bm\Delta=\RR'-\RR$,
\begin{align}
  \ee_{\RR\mu}^{\mathsf T}\mathcal{J}_{\mu\nu}(\bm\Delta)\ee_{\RR'\nu}
  &=({\bf m}_{\RR\mu}+\delta\ee_{\RR\mu})^{\mathsf T}
    \mathcal{J}_{\mu\nu}(\bm\Delta)
    ({\bf m}_{\RR'\nu}+\delta\ee_{\RR'\nu}) \\
  &=
    {\bf m}_{\RR\mu}^{\mathsf T}\mathcal{J}_{\mu\nu}(\bm\Delta){\bf m}_{\RR'\nu}
    +\delta\ee_{\RR\mu}^{\mathsf T}\mathcal{J}_{\mu\nu}(\bm\Delta){\bf m}_{\RR'\nu}
    +{\bf m}_{\RR\mu}^{\mathsf T}\mathcal{J}_{\mu\nu}(\bm\Delta)\delta\ee_{\RR'\nu}
    +\delta\ee_{\RR\mu}^{\mathsf T}\mathcal{J}_{\mu\nu}(\bm\Delta)\delta\ee_{\RR'\nu} .
\end{align}
In the mean-field approximation, the last term, which contains correlations between fluctuations on different sites, is neglected.
Therefore,
\begin{equation}
  \ee_{\RR\mu}^{\mathsf T}\mathcal{J}_{\mu\nu}(\bm\Delta)\ee_{\RR'\nu}
  \simeq
  \ee_{\RR\mu}^{\mathsf T}\mathcal{J}_{\mu\nu}(\bm\Delta){\bf m}_{\RR'\nu}
  +
  {\bf m}_{\RR\mu}^{\mathsf T}\mathcal{J}_{\mu\nu}(\bm\Delta)\ee_{\RR'\nu}
  -
  {\bf m}_{\RR\mu}^{\mathsf T}\mathcal{J}_{\mu\nu}(\bm\Delta){\bf m}_{\RR'\nu} .
  \label{eq:mf_pair}
\end{equation}
Using the symmetry condition in Eq.~\eqref{eq:reality}, the two linear terms give identical contributions after summation over all pairs. Thus the mean-field Hamiltonian becomes
\begin{equation}
  E^{\rm MF}
  =
  -2\sum_{\RR\mu}
  \bm h_{\RR\mu}^{\rm MF}\cdot\ee_{\RR\mu}
  +
  \sum_{\RR,\RR'}\sum_{\mu\nu}
  {\bf m}_{\RR\mu}^{\mathsf T}\mathcal{J}_{\mu\nu}(\RR'-\RR){\bf m}_{\RR'\nu},
  \label{eq:H_MF}
\end{equation}
where
\begin{equation}
  \bm h_{\RR\mu}^{\rm MF}
  =
  \sum_{\RR'\nu}
  \mathcal{J}_{\mu\nu}(\RR'-\RR){\bf m}_{\RR'\nu} .
  \label{eq:h_def}
\end{equation}
The single-site partition function is given by
\begin{equation}
  Z_{\RR\mu} = \int \dif\Omega_{\RR\mu} \exp\left( 2\beta \bm h_{\RR\mu}^{\rm MF
  }\cdot\ee_{\RR\mu} \right) = \frac{2\pi}{\beta h_{\RR\mu}^{\rm MF}} \sinh(2\beta h_{\RR\mu}^{\rm MF}).
\end{equation}
This gives
\begin{align}
  \begin{split}
  {\bf m}_{\RR \mu} &= \frac{1}{Z_{\RR\mu}} \int \dif\Omega_{\RR\mu} \ee_{\RR\mu} \exp\left( 2\beta \bm h_{\RR\mu}^{\rm MF}\cdot\ee_{\RR\mu} \right)\\
  &= \left( \coth(2\beta h_{\RR\mu}^{\rm MF}) - \frac{1}{2\beta h_{\RR\mu}^{\rm MF}} \right) \hat{\bm h}_{\RR\mu}^{\rm MF}, \qquad \hat{\bm h}_{\RR\mu}^{\rm MF} = \frac{\bm h_{\RR\mu}^{\rm MF}}{h_{\RR\mu}^{\rm MF}}.
  \label{eq:m_self}
  \end{split}
\end{align}
Note that Eq.~\eqref{eq:m_self} is a self-consistent equation for the order parameter ${\bf m}_{\RR\mu}$, since $\bm h_{\RR\mu}^{\rm MF}$ depends on ${\bf m}_{\RR'\nu}$ via Eq.~\eqref{eq:h_def}.

Near the magnetic transition temperature, the magnitude of the order parameter is sufficiently small.
Therefore, Eq.~\eqref{eq:m_self} can be expanded with respect to
\begin{equation}
  x_{\RR\mu}=2\beta h_{\RR\mu}^{\rm MF}.
\end{equation}
Using the Taylor expansion of the Langevin function,
\begin{equation}
  L(x)
  =
  \coth x - \frac{1}{x}
  =
  \frac{x}{3}
  -
  \frac{x^3}{45}
  +
  O(x^5),
  \label{eq:Langevin_expansion}
\end{equation}
we obtain, to the leading order,
\begin{equation}
  {\bf m}_{\RR\mu}
  =
  \frac{2\beta}{3}
  \bm h_{\RR\mu}^{\rm MF}
  +
  O(|{\bf m}|^3).
  \label{eq:m_linearized}
\end{equation}
Substituting Eq.~\eqref{eq:h_def} into Eq.~\eqref{eq:m_linearized} gives the linearized mean-field equation
\begin{equation}
  {\bf m}_{\RR\mu}
  =
  \frac{2\beta}{3}
  \sum_{\RR'\nu}
  \mathcal{J}_{\mu\nu}(\RR'-\RR)
  {\bf m}_{\RR'\nu}.
  \label{eq:linearized_real}
\end{equation}
The lattice Fourier transform of the order parameter is defined as
\begin{equation}
  {\bf m}_{\mu}(\qv) = \frac{1}{\sqrt{N}} \sum_{\RR} e^{-i\qv\cdot\RR} {\bf m}_{\RR\mu},
  \label{eq:fourier_m}
\end{equation}
where $N$ is the number of Bravais lattice cells.
The exchange coupling tensor in Fourier space is defined as
\begin{equation}
  \mathcal{J}_{\mu\nu}(\qv)
  =
  \sum_{\bm\Delta}
  e^{+i\qv\cdot\bm\Delta}
  \mathcal{J}_{\mu\nu}(\bm\Delta).
  \label{eq:fourier_J}
\end{equation}
Then, Eq.~\eqref{eq:linearized_real} can be rewritten in the Fourier space as
\begin{equation}
  {\bf m}_{\mu}(\qv)
  =
  \frac{2\beta}{3}
  \sum_{\nu}
  \mathcal{J}_{\mu\nu}(\qv)
  {\bf m}_{\nu}(\qv).
  \label{eq:linearized_fourier}
\end{equation}
Collecting the spin and sublattice components into a $3n_{\rm sub}$-dimensional vector,
\begin{equation}
  {\bf M}(\qv) = \begin{pmatrix}
  {\bf m}_1(\qv) \\
  {\bf m}_2(\qv) \\
  \vdots \\
  {\bf m}_{n_{\rm sub}}(\qv)
  \end{pmatrix},
\end{equation}
we can rewrite Eq.~\eqref{eq:linearized_fourier} as
\begin{equation}
  {\bf M}(\qv) = \frac{2\beta}{3}\mathcal{J}(\qv) {\bf M}(\qv),
\end{equation}
where $\mathcal{J}(\qv)$ is the $3n_{\rm sub}\times 3n_{\rm sub}$ matrix defined by
\begin{equation}
  \mathcal{J}(\qv) = \begin{pmatrix}
  \mathcal{J}_{11}(\qv) & \mathcal{J}_{12}(\qv) & \cdots & \mathcal{J}_{1n_{\rm sub}}(\qv) \\
  \mathcal{J}_{21}(\qv) & \mathcal{J}_{22}(\qv) & \cdots & \mathcal{J}_{2n_{\rm sub}}(\qv) \\
  \vdots & \vdots & \ddots & \vdots \\
  \mathcal{J}_{n_{\rm sub}1}(\qv) & \mathcal{J}_{n_{\rm sub}2}(\qv) & \cdots & \mathcal{J}_{n_{\rm sub}n_{\rm sub}}(\qv)
  \end{pmatrix}.
\end{equation}
From Eq.~\eqref{eq:reality}, $\mathcal{J}(\qv)$ is Hermitian, and its eigenvalues are real. A nonzero solution at $\qv$ appears when
\begin{equation}
  \frac{2\beta}{3} \eta_{n}(\mathcal{J}(\qv)) = 1,
\end{equation}
where $\eta_{n}(\mathcal{J}(\qv))$ is the $n$-th eigenvalue of $\mathcal{J}(\qv)$.
Therefore, the first instability from the paramagnetic state occurs for the mode that maximizes the largest eigenvalue of $\mathcal{J}(\qv)$:
\begin{equation}
  \qv_{\rm max} = \arg\max_{\qv} \eta_{\rm max}(\mathcal{J}(\qv)),
\end{equation}
and the spin-spiral period is given by $\lambda = 2\pi/|\qv_{\rm max}|$.

\begin{figure}[h]
  \includegraphics[width=120mm]{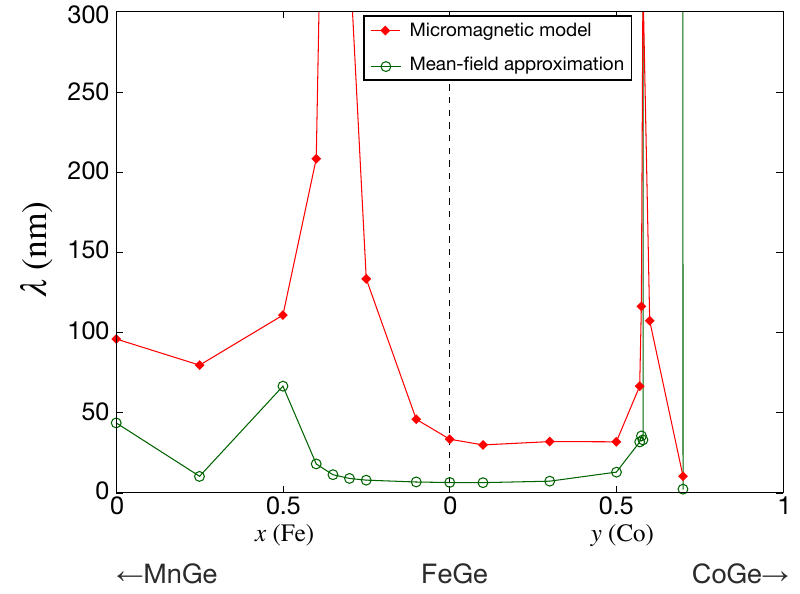}
  \caption{Spin-spiral period $\lambda$ evaluated for the spin model derived in the main text using the mean-field approximation (green circles) and the micromagnetic model (red diamonds) for ${\rm Mn}_{1-x}{\rm Fe}_{x}{\rm Ge}$ and ${\rm Fe}_{1-y}{\rm Co}_{y}{\rm Ge}$.}  
\label{fig:lambda_mf}
\end{figure}

Figure~\ref{fig:lambda_mf} shows the spin-spiral period $\lambda$ in ${\rm Mn}_{1-x}{\rm Fe}_{x}{\rm Ge}$ and ${\rm Fe}_{1-y}{\rm Co}_{y}{\rm Ge}$, evaluated using the mean-field approximation and the micromagnetic model.
In both cases, the analysis is based on the same spin model derived in the main text.
Although the absolute values of $\lambda$ differ between the two approaches, their overall composition dependences are similar.
This result supports the conclusion that the spin model derived in this work stabilizes a spin-spiral state.

